\documentclass{aa}

\usepackage{array}
\usepackage{amsmath}
\usepackage{graphicx}
\usepackage{txfonts}

\graphicspath{{./}{figures/}}

\begin{document}

\linenumbers
\renewcommand\makeLineNumber{}

\title{Solar wind data analysis aided by synthetic modeling: a better understanding of plasma-frame variations from temporal data}

   \author{N. Magyar
          \inst{1} \fnmsep\thanks{FWO postdoctoral fellow;               \email{norbert.magyar@kuleuven.be}}
          \and
           J. Verniero
           \inst{2}
           \and 
           A. Szabo
           \inst{2}
           \and 
           J. Zhang
           \inst{3}
           \and
           T. Van Doorsselaere
           \inst{1}
           }

   \institute{Centre for mathematical Plasma Astrophysics (CmPA), Department of Mathematics, 
KU Leuven, Celestijnenlaan 200B bus 2400, \\ B-3001 Leuven, Belgium\\
   \and
    Code 672, NASA, Goddard Space Flight Center, Greenbelt, MD 20771, USA\\
    \and
    Section of Analysis, Department of Mathematics,
KU Leuven, Celestijnenlaan 200B bus 2400, \\ B-3001 Leuven, Belgium
    }

\abstract
 {In-situ measurements of the solar wind, a turbulent and anisotropic plasma flow originating at the Sun, are mostly carried out by single spacecraft, resulting in one-dimensional time series.}
 {The conversion of these measurements to the spatial frame of the plasma is a great challenge, but required for direct comparison of the measurements with MHD turbulence theories.} 
 {Here we present a toolkit, based on the synthetic modeling of solar wind fluctuations as two-dimensional noise maps with adjustable spectral and power anisotropy, that can help with the temporal-spatial conversion of real data. Specifically, by following the spacecraft trajectory through a noise map (relative velocity and angle relative to some mean magnetic field) with properties tuned to mimic those of the solar wind, the likelihood that the temporal data fluctuations represent parallel or perpendicular fluctuations in the plasma frame can be quantified by correlating structure functions of the noise map. Synthetic temporal data can also be generated, which can provide a testing ground for analysis applied to the solar wind data.}
 {We demonstrate this tool by investigating Parker Solar Probe's E7 trajectory and data, and showcase several possible ways in which it can be used. We find that whether temporal variations in the spacecraft frame come from parallel or perpendicular variations in the plasma frame strongly depends on the spectral and power anisotropy of the measured wind.} 
 {Data analysis assisted by such underlying synthetic models as presented here could open up new ways to interpret measurements in the future, specifically in the more reliable determination of plasma frame quantities from temporal measurements.}
 {}

\maketitle 

\section{Introduction} \label{sec:intro}

The solar wind is a continuous supersonic and super-Alfv\'enic outflow of ionized particles (mostly protons and electrons) from the Sun, driven partly by an as-of-yet undiscovered coronal heating mechanism \citep{2006SoPh..234...41K,2015RSPTA.37340261A,2020SSRv..216..140V} which renders the multi-million K solar corona unable to form (hydro)static equilibrium, as first shown by \citet{1958ApJ...128..664P}. Since the first experimental evidence of the solar wind provided by early satellites \citep{1960SPhD....5..361G,1962Sci...138.1095N}, much effort was devoted to understanding the measured fluctuations in velocity, magnetic field, density, among other properties. It was soon realized that solar wind fluctuations display both turbulent \citep{1968ApJ...153..371C} and Alfvén wave \citep{1971JGR....76.3534B} characteristics on a large range of scales, the duality of which is thought to be the basis for understanding the nature of solar wind fluctuations, completed by the presence of advected non-Alfvénic and large-scale structures \citep[][although see \citealt{10.3389/fspas.2020.00020} for a different viewpoint]{1995SSRv...73....1T,2013LRSP...10....2B}. An important property of solar wind fluctuations is their anisotropy with respect to the local magnetic field direction, such as spectral, power, variance, and wavevector anisotropies \citep{2012SSRv..172..325H}. Moreover, perpendicular fluctuations usually lack axisymmetry around the local magnetic field direction. \par 
The interpretation of in-situ data is complicated by the fact that most solar wind measurements rely on data from a single satellite or probe, that is, they represent single-point measurements along the relative plasma frame\footnote{`Plasma frame' usually means a frame with no mean bulk flow, that is, the one `carried' by the background solar wind.} trajectory of the spacecraft. Therefore, the analysis of the provided one-dimensional temporal data requires a set of assumptions to be made for its re-interpretation as spatial data in the reference frame of the wind. This step is necessary in order to compare measurements to theoretical results which are given as functions of spatial variables or correlations. An almost universally applied assumption is Taylor's hypothesis \citep{1938RSPSA.164..476T}, which states that temporal variations in the spacecraft frame simply equal spatial variations in the plasma frame, $t = -\mathbf{V}_{SW} x$, where $\mathbf{V}_{SW}$ is the solar wind speed vector. This can often be justified simply based on the observation that wind speeds are usually much faster than the characteristic velocity fluctuations and wave speeds in the plasma rest frame, rendering the fluctuations to appear `frozen in' the plasma as they are advected past the spacecraft. 
Parker Solar Probe (PSP), launched in 2018, has become the first satellite to enter the magnetically dominated solar corona, nearly three years after its launch \citep{2021PhRvL.127y5101K}, and continues to gather magnetic field and particle data, among other measurements, on successive encounters with perihelia as close to the solar surface as 10 $R_\odot$. Recently, a considerable amount of work has been dedicated to studying the validity of Taylor's hypothesis under various conditions, with the anticipation of PSP measurements for which the above inequality might not hold \citep[e.g.,][]{2014ApJ...790L..20K,2017AnGeo..35..325N,2018ApJ...858L..20B,2019ApJS..242...12C,2021A&A...650A..22P}, based on various phenomenological, analytical or numerical models of solar wind turbulence. Uncertainties related to the interpretation of temporal data from PSP as spatial variations constitutes a motivation for applying the current analysis specifically to PSP data. \par 
In studies of solar wind anisotropy, it is often assumed that for spacecraft trajectories in the plasma rest frame at shallow angles with respect to the local mean magnetic field (e.g. within 15\textdegree), the measurements sample parallel ($ k_\parallel$) variations \citep{2019ApJ...887..160T}. Various methods have been applied to solar wind measurements in order to reveal the anisotropy of the fluctuations as a function of the `angle of attack' of the spacecraft with respect to the local mean magnetic field. The spectral anisotropy of the solar wind fluctuation has been studied using wavelet analysis \citep[e.g., by ][to name a few]{2008PhRvL.101q5005H,2009ApJ...698..986P,2011MNRAS.415.3219C,2016ApJ...816...15W,2022MNRAS.514.1282S}, or Hilbert spectral analysis \citep{2019ApJ...887..160T}, among other methods \citep{2015ApJ...810L..21W,2022ApJ...933...56A,2022ApJ...924L...5Z} as a function of various conditions, such as intermittency, distance from the Sun, presence of switchbacks, etc. Despite these large numbers of studies, there is still no consensus on the spectral anisotropy of the solar wind, even on the same data set, illustrating the aforementioned difficulty in transforming temporal measurements into plasma frame data. Determining spectral anisotropy is important, for instance, because some theoretical works \citep{1995ApJ...438..763G} postulate the existence of critical balance in MHD turbulence, a manifestation of which are specific spectral power slopes along ($E(k_\parallel) \sim k_\parallel^{-2}$) and across ($E(k_\perp) \sim k_\perp^{-5/3}$) the local mean magnetic field, respectively. It is clear that a better knowledge of spatial variations of fluctuations would greatly advance our overall understanding of solar wind dynamics. For instance, a reliable determination of linear and nonlinear advective terms of the MHD equations in the solar wind could validate theories and phenomenologies of turbulence generation and evolution.  
\par
In this paper, we aim to assess the validity of interpreting temporal data as spatial plasma frame data through single-point solar wind measurements by developing a synthetic model of solar wind fluctuations. We consider the 7\textsuperscript{th} encounter of PSP as the spacecraft trajectory through the synthetic model, with spacecraft speeds relative to the plasma frame and angles with respect to the local mean field given by PSP measurements. This allows for the direct comparison of the 1D synthetic temporal data in the spacecraft frame to the underlying synthetic spatial data. Among other tests, we explore the conditions under which the parallel or perpendicular fluctuations can be sampled depending on the degree of anisotropy, test the limits of Taylor's hypothesis, and estimate the reliability in determining spatial gradients from temporal derivatives.
The paper is organized as follows. In Sect.~\ref{sec:model}, we describe in detail the synthetic model employed. In Sect.~\ref{sec:data} we describe the PSP data for the chosen encounter. In Sect.~\ref{sec:results}, we present several comparisons between the temporal and spatial data based on parametric studies of the synthetic model. Finally, in Sect.\ref{sec:concl}, we conclude the results and discuss possible future applications and updates of the presented model. 

\section{Synthetic Model} \label{sec:model}

\begin{figure*}
    \centering
       \begin{tabular}{@{}cc@{}}
        \includegraphics[width=0.4\textwidth]{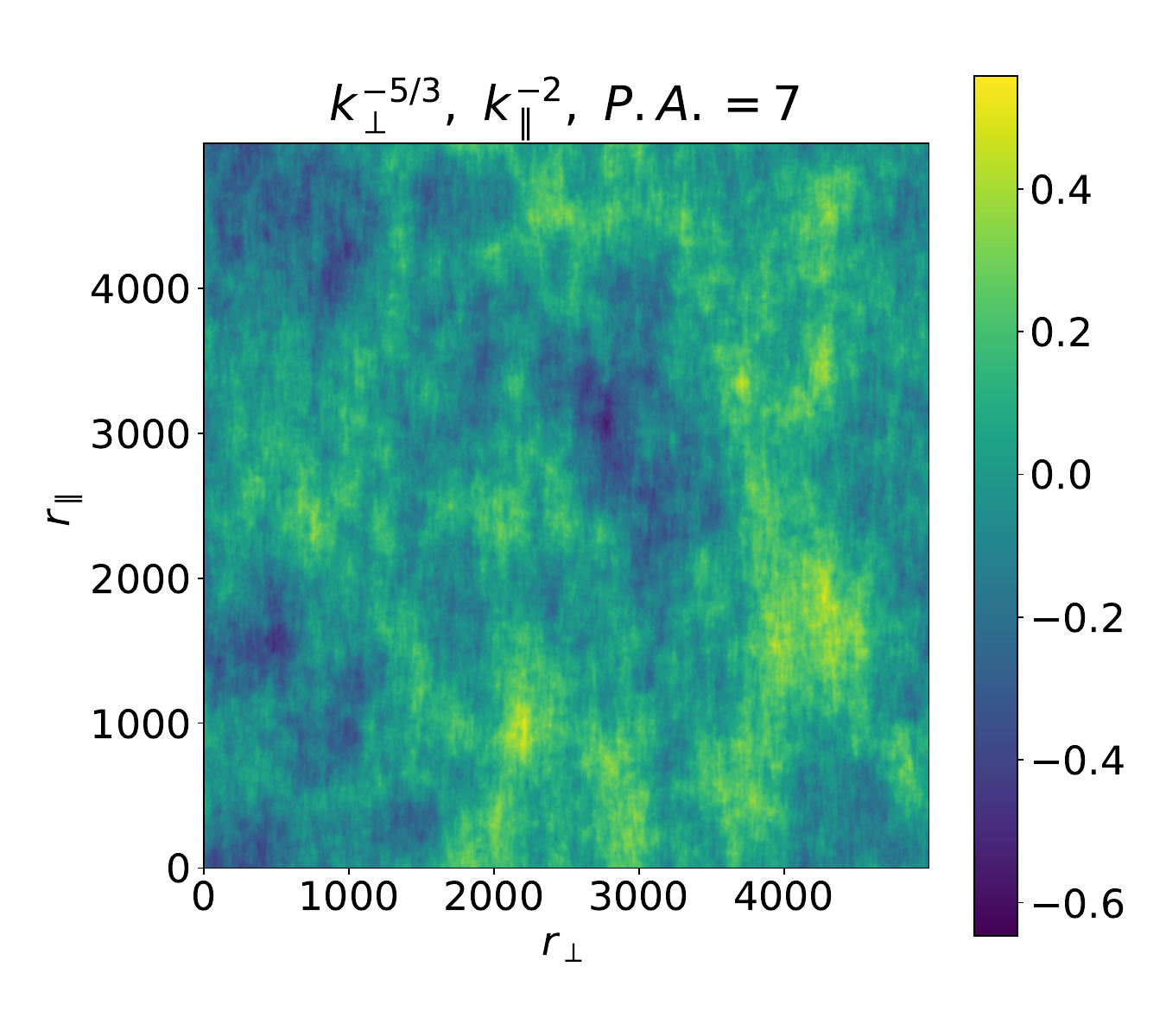}  
        \includegraphics[width=0.35\textwidth]{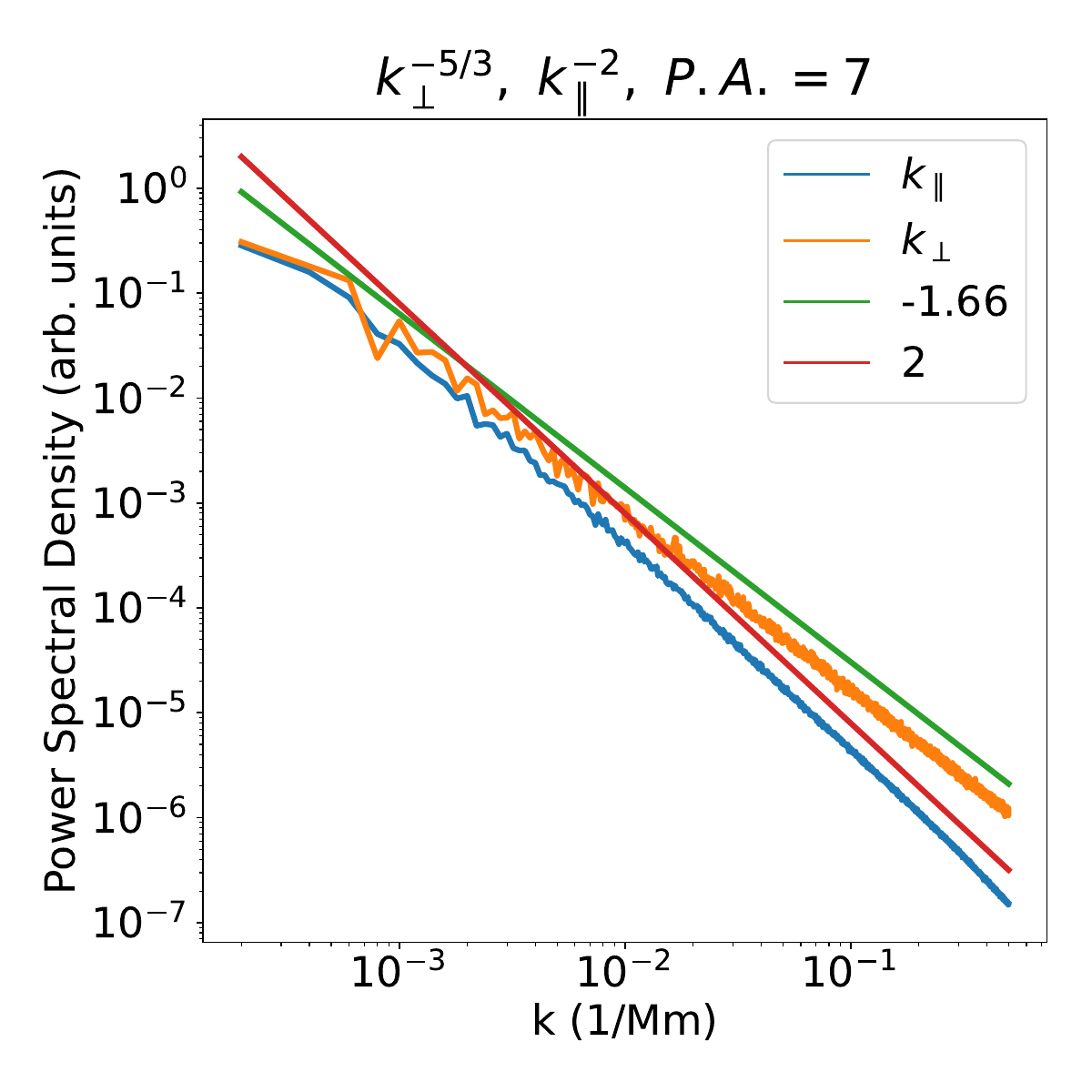} \\
       \end{tabular}  
        \caption{Anisotropic 2D noise map (\textit{left}), and its 1D Fourier power spectra averaged over all rows and columns, respectively (\textit{right}). P.A. means `Power Anisotropy', that is, the ratio of the power in the perpendicular to parallel wave number at the highest wave number. $r_\perp$ and $r_\parallel$ are the array indices in the noise map perpendicular and parallel to the magnetic field, respectively. The amplitude of the noise map is shown in arbitrary units.}
        \label{figure01}
 \end{figure*}

\begin{figure*}
    \centering
       \begin{tabular}{@{}ccc@{}}
        \includegraphics[width=0.33\textwidth]{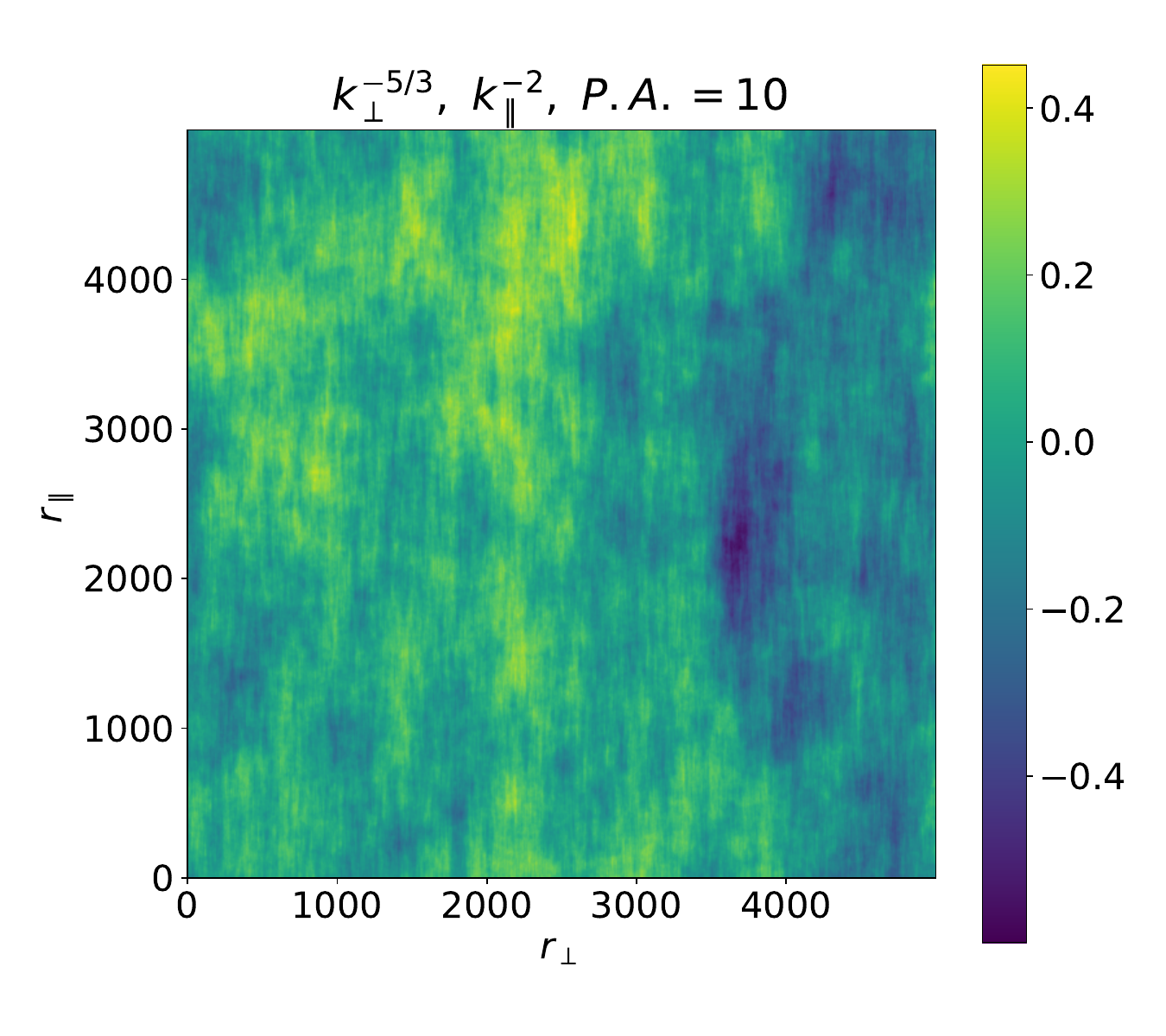}  
        \includegraphics[width=0.33\textwidth]{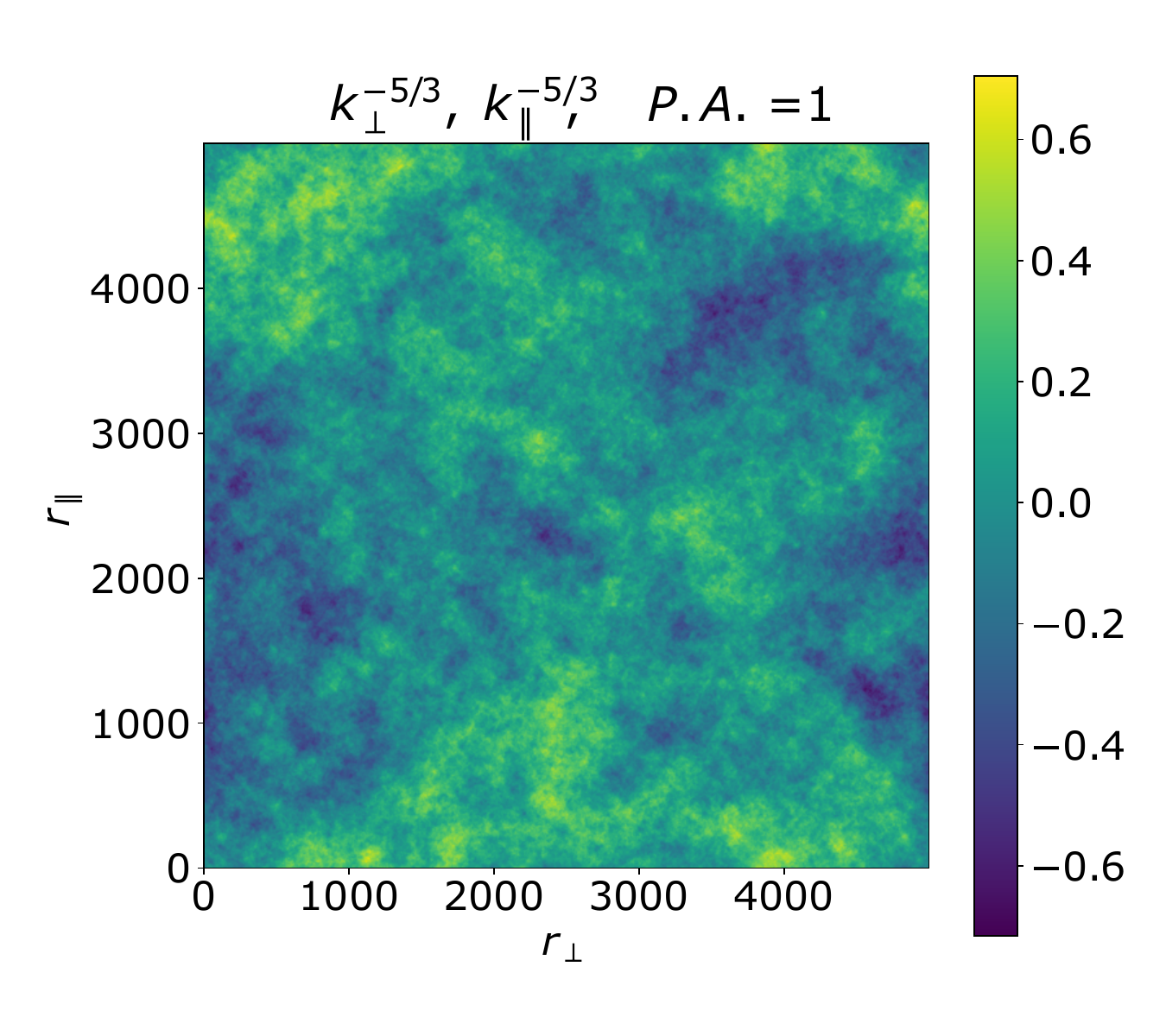}  
        \includegraphics[width=0.33\textwidth]{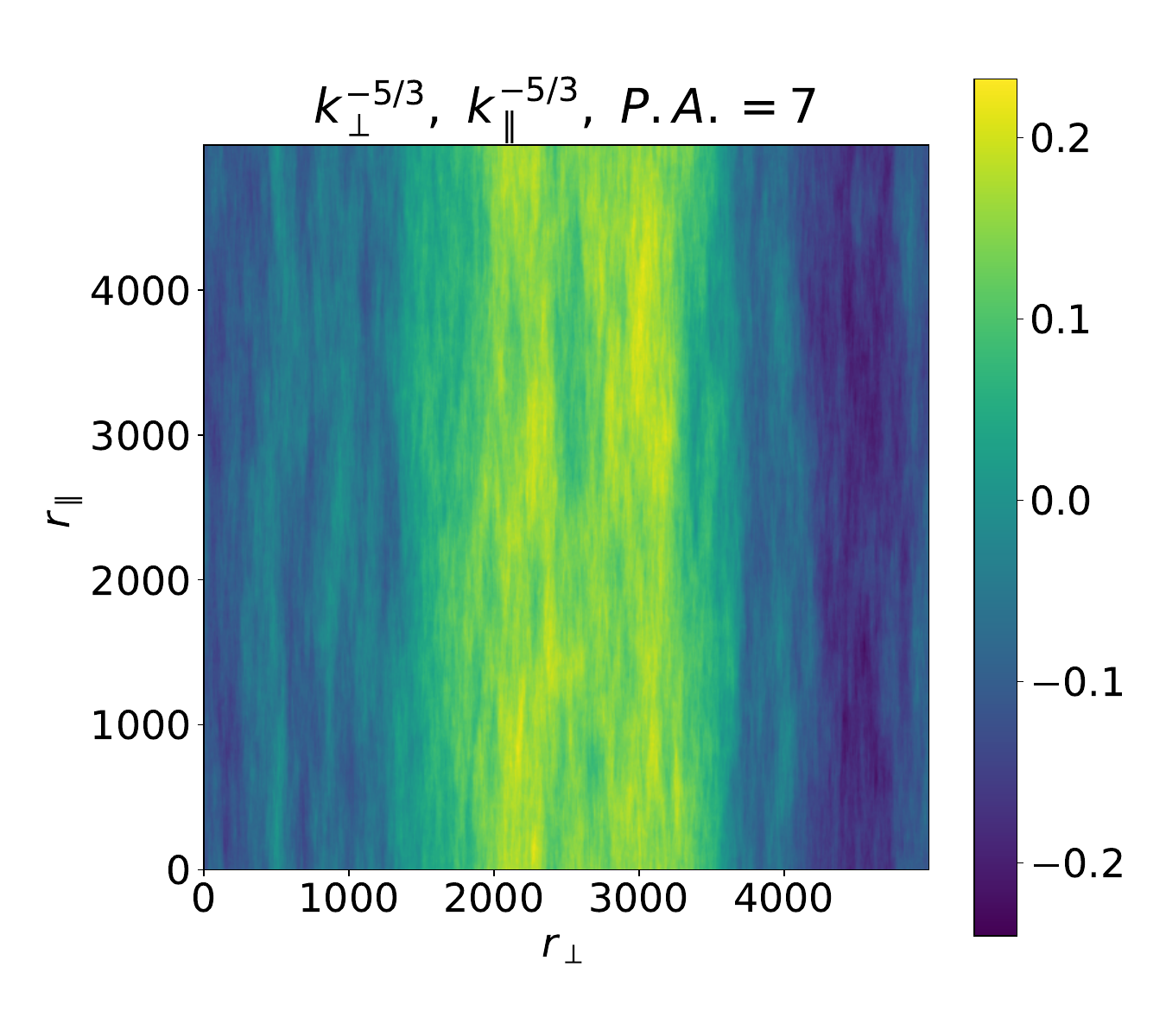} 
       \end{tabular}  
        \caption{Noise maps of different spectral and power anisotropies (P.A.), shown in the title of each map. The fluctuation amplitudes are in arbitrary units.}
        \label{figure02}
 \end{figure*}

For the synthetic solar wind fluctuations, two-dimensional (2D) arrays are constructed, using the \texttt{numpy} package \citep{harris2020array} in \texttt{Python}. The choice of 2D arrays instead of 3D ones is made, on the one hand, based on the fact that the PSP orbit is sampling plasma in Venus' orbital plane within 4\textdegree\ inclination to the Sun's equator, with negligible mean solar wind flow normal to this plane, and on the other hand, for computational cost reduction. The fluctuations are constructed in the following way. An $N^2$ 2D array is populated with white noise (flat power spectrum) using the \texttt{random.randn} routine, which returns random floats sampled from a univariate normal (Gaussian) distribution. A real-valued filter is applied in $k$     -space (element-wise multiplication) to the 2D FFT of the white noise array, which is an $N^2$ 2D array that determines the desired amplitudes and power spectral slopes for the $k_\parallel$ and $k_\perp$ directions, which are aligned with the principal axes of the matrix. There are multiple ways to construct such a filter. With the zero wavenumber at $(N/2,N/2)$, that is, center-shifted spectrum, the positive-valued $k_\parallel, k_\perp$ increase linearly towards the edges of the 2D array, up to the Nyquist frequency, symmetric across both principal axes. Here we use N = 5000. 
The problem of finding a 2D filter with specified average power laws in the two directions is almost equivalent to the problem in probability theory of finding a joint distribution given two marginal distributions, where an integrability condition is further required. In the latter case, the standard procedure is to choose a copula function, which is the joint probability distribution with its two marginal distributions being the constant function on the interval $[0,1]$, and generate the desired joint distribution by taking inverse functions on the marginals. The simplest copula is the product copula, which assumes that the joint distribution is only the product of the two marginal distributions. In this case the filter has the following form: : 
\begin{equation}
    F(k_\perp,k_\parallel) = 
    \begin{cases}
     A/k_\perp^\alpha\ & \text{if}\ k_\perp \neq 0 ,\ k_\parallel = 0,\\
     B/k_\parallel^\beta\ & \text{if}\ k_\perp = 0,\ k_\parallel \neq 0,\\
     A B/(k_\perp^\alpha k_\parallel^\beta)\ & \text{if}\ k_\perp \neq 0,\ k_\parallel \neq 0,
    \end{cases}
    \label{filter1}
\end{equation}
The product copula is highly anisotropic, having much less power in the oblique Fourier modes, resulting in very pronounced vertical and horizontal lines in real space. Since fluctuations are mostly spherically polarized in the solar wind \citep[e.g.,][]{1998JGR...103..335V}, it may be more realistic to consider a smoother (more isotropic) variation between the parallel and perpendicular directions, while still resulting in the desired averaged spectra along these directions. While more sophisticated copulas exist to yield such a result, they have more involved functional forms with tens of terms including special functions, which may not be the most convenient choice for an application in our problem. Instead, we have experimented with filters of the following form:
\begin{equation}
    F(k_\perp,k_\parallel) = \frac{1}{(A  k_\perp^\alpha + B k_\parallel^\beta)^{1/2}} + \frac{C}{(k_\perp^2 + k_\parallel^2)^{\gamma/2}},
    \label{filter2}
\end{equation}
where $A,B,C, \alpha,\beta,\gamma$ are constants that define the resulting 2D power spectrum. With $F$ applied, we perform an inverse Fourier transform of the 2D array, resulting in an anisotropic colored noise 2D array, which we refer to in the following as the noise map. The power spectrum in the parallel and perpendicular directions is given by averaging 1D power spectra over all rows and columns of the noise map, respectively. This method of calculating parallel and perpendicular spectra is consistent with how these are calculated for real data. Natural choices for parallel and perpendicular 1D power spectra and their ratios are the ones that were determined from in-situ observations in previous works. Various studies of noise maps with different spectral and power anisotropies will be presented in the Results section. Spectral anisotropy refers to different slope values for parallel and perpendicular power spectra, while power anisotropy refers to the ratio between perpendicular and parallel power ($P_\perp/ P_\parallel$) at some specific scale, here chosen to be at the Nyquist frequency, that is, $f$ = 0.5/cadence. If the spectral anisotropy is known, the power anisotropy of the noise maps can be calculated at any other scale based on this value. An example of a noise map and its parallel and perpendicular power spectra is illustrated in Fig.~\ref{figure01}.
Examples of some of the other noise maps used in this study are presented in Fig.~\ref{figure02}. 
The constants used for noise maps with different power and spectral anisotropies are shown in Table~\ref{table1}.
\begin{table}[h!]
\centering
\resizebox{\columnwidth}{!}{%
\begin{tabular}{||c|c||c|c|c|c|c|c||} 
\hline
P.A. & Spectra & A & B & C & $\alpha$ & $\beta$ & $\gamma$ \\
\hline
1 & $k_\perp^{-5/3},\ k_\parallel^{-5/3}$ & 1 & 1 & 20 & 2.66 & 2.66  & 0.1\\ 
7 & $k_\perp^{-5/3},\ k_\parallel^{-5/3}$ & 0.01 & $3.5 \cdot 10^{-6}$ & 6000 & 2.75 & 2.75 & 1  \\ 
5 & $k_\perp^{-5/3},\ k_\parallel^{-2}$ & 100 & $6 \cdot 10^{-4}$ & 400 & 4 & 1.9 & 1.4  \\ 
7 & $k_\perp^{-5/3},\ k_\parallel^{-2}$ & 100 & $10^{-3}$ & 200 & 3.65 & 2.13 & 1.4  \\ 
10 & $k_\perp^{-5/3},\ k_\parallel^{-2}$ & 100 & $5 \cdot 10^{-4}$  & 200 & 3.4 & 2.25& 1.4 \\
\hline
\end{tabular}
}
\caption{The constants used in Eq.~\ref{filter2} for noise maps with different power (P.A.) and spectral anisotropies}
\label{table1}
\end{table}

\section{Data Description} \label{sec:data}

\begin{figure*}
    \centering
    \includegraphics[width=1.0\textwidth]{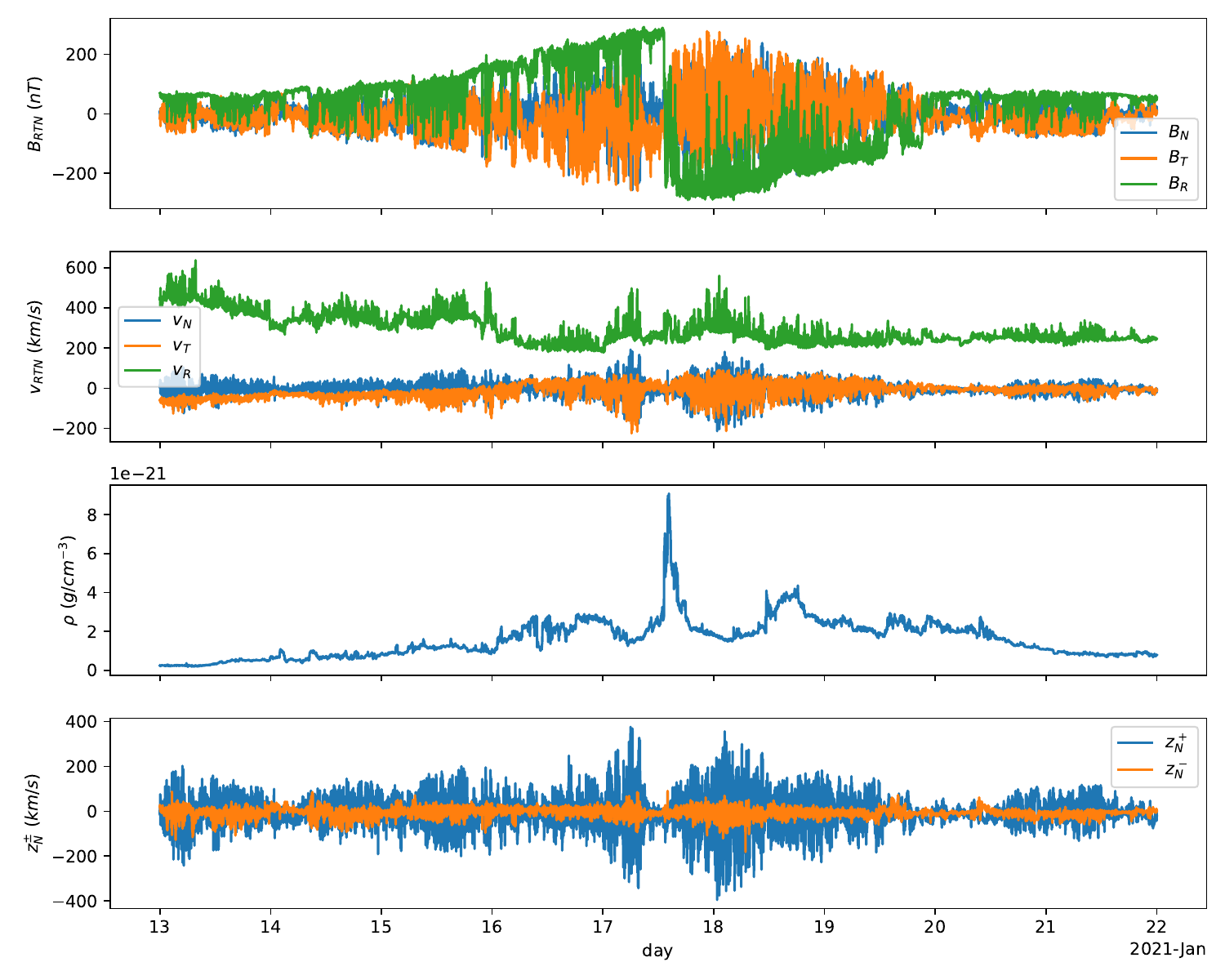}
    \caption{In-situ data from PSP, showing in RTN coordinates the magnetic field, velocity, density, and normal component of Els\"{a}sser variables, from top to bottom rows respectively, during the analyzed time interval.}
    \label{figure1}
\end{figure*}

\begin{figure*}
    \centering
    \includegraphics[width=0.5\textwidth]{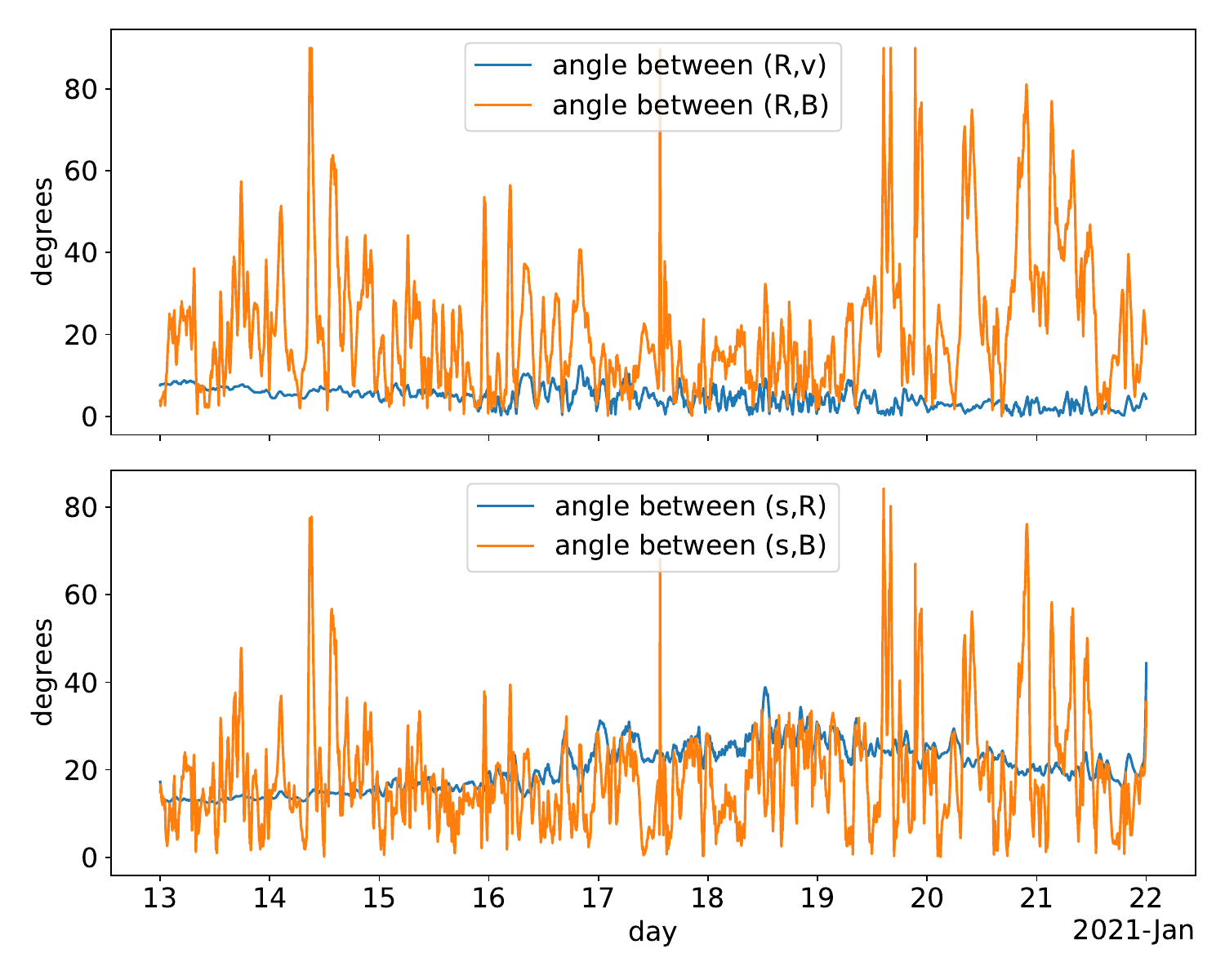}
    \caption{Plots of the angle between the radial direction and 30 min moving average velocity and magnetic field direction (top graph), and of the angle between the PSP velocity vector relative to the background flow (s) and the radial direction and average magnetic field direction (bottom graph), for the in-situ PSP data shown in Fig.~\ref{figure1}}
    \label{figure102}
\end{figure*}

We analyze data from the 7\textsuperscript{th} encounter of PSP, from Jan 13 2021 00:00UTC to Jan 22 2021 00:00UTC. The data is publicly available from CDAWeb\footnote{cdaweb.gsfc.nasa.gov/}. From the FIELDS instrument suite \citep{2016SSRv..204...49B}, we use 3D fluxgate magnetometer data at 4 vectors/s in RTN coordinates, `mag\_rtn\_4\_per\_cycle', and electron density measured through quasi-thermal noise. From the SWEAP \citep{2016SSRv..204..131K} instrument suite, we use proton data from the SPAN-I instrument \citep{2022ApJ...938..138L} for velocity perturbations. We interpolate magnetic and density data to have the same cadence as velocity data, $dt = 3.49\ $ s. We apply the Butterworth filter (\texttt{scipy.signal.butter} of order 5 and critical frequency 0.03 in units of cadence) to the density data $\rho$ in order to smooth its characteristic step-like instrumental oscillation without affecting the overall trend. The magnetic field data $B$ is converted to velocity units $b = B \rho^{-0.5}$. Average values of velocity and magnetic field in velocity units are calculated by means of moving average, with a window size of 30 minutes. The \citet{1950PhRv...79..183E} variables are calculated as:
\begin{equation}
    \mathbf{z}^\pm = \mathbf{v} \mp sign(B_{0r}) \mathbf{b},
\end{equation}
where $sign(B_{0r})$ is the sign of the average radial magnetic field, assuring that the dominant, outward-propagating Els\"{a}sser variable is $z^+$, irrespective of the background magnetic field polarization. 
The full data for the analyzed duration is shown in Fig.~\ref{figure1}.
Next, we proceed to calculate the angle between various vectors, such as the radial direction, average magnetic field, and relative velocity, given by the arc cosine of the dot product between the respective normalized vectors. The results are plotted in Fig.~\ref{figure102}. 
The true `angle of attack' to the local mean magnetic field in the spacecraft frame is given by the angle between the relative velocity vector ($V_{SC} - V_{SW}$) and local average magnetic field direction. Note in Fig.~\ref{figure102} that most of the angle of attack variation is given by the variation of the average magnetic field direction, superimposed on the weaker and longer trend represented by the angle between the relative spacecraft speed and the radial direction. 

\section{Results} \label{sec:results}

\subsection{Gradient correlations of noise maps}

\begin{figure*}
    \centering
       \begin{tabular}{@{}ccc@{}} 
        \includegraphics[width=0.33\textwidth]{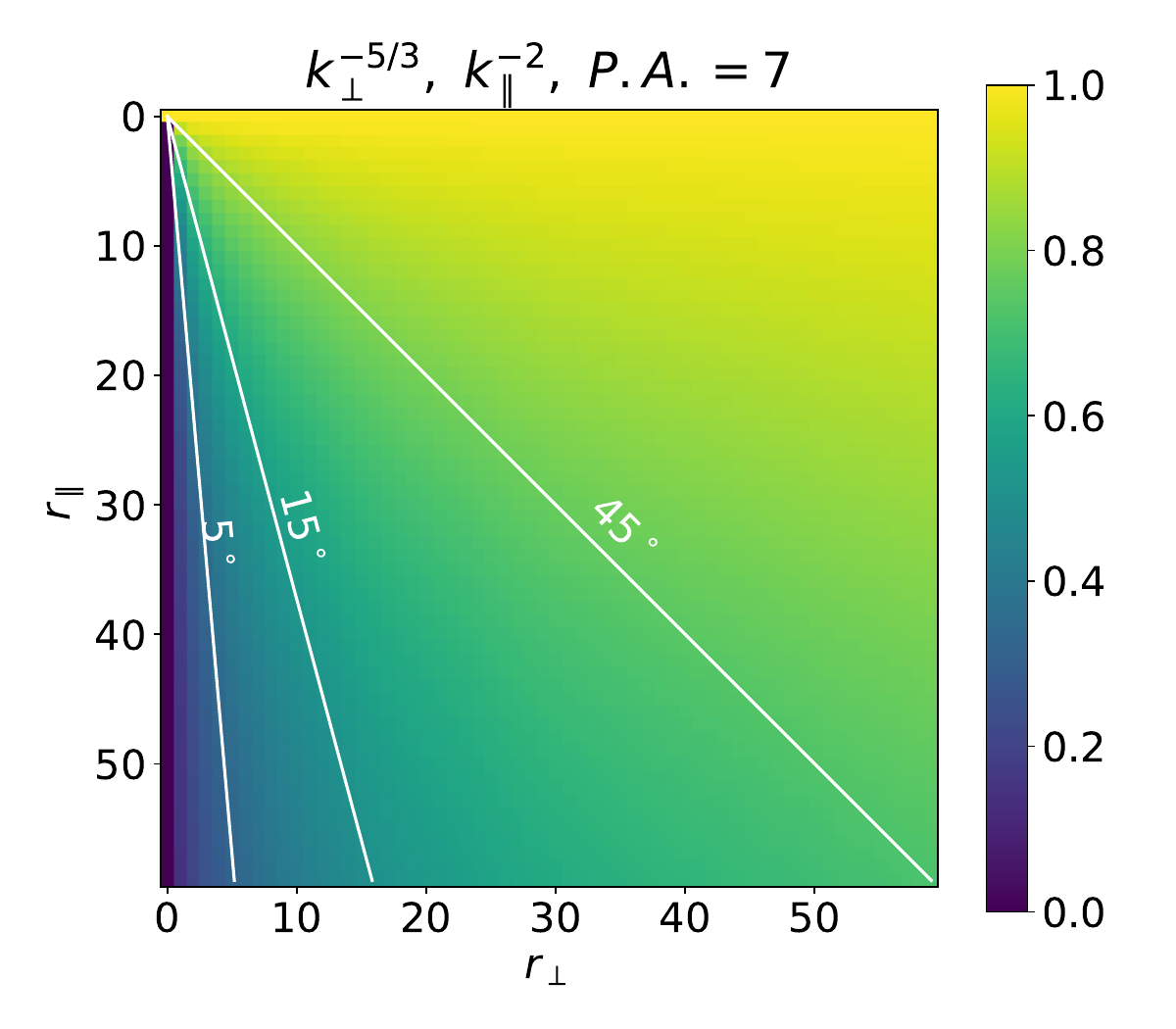}
        \includegraphics[width=0.33\textwidth]{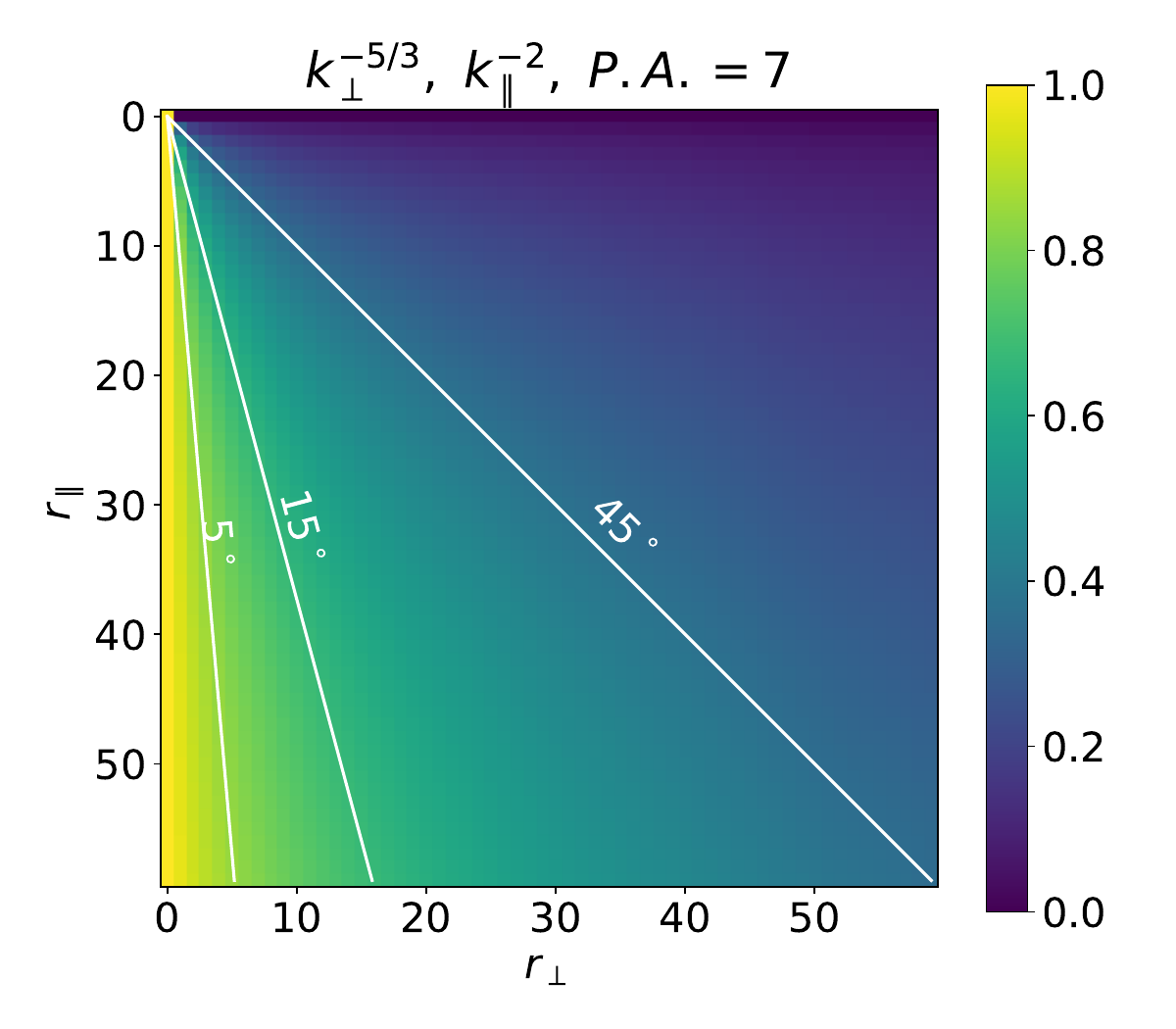} \\
       \end{tabular}  
        \caption{Cross-correlation map of total ($dt$) and perpendicular (left, $\partial T$) or parallel (right, $\partial R$) gradients, respectively, for different $\mathbf{r}$, ranging from 0 to 60 in both perpendicular ($r_\perp$) and parallel ($r_\parallel$) directions. Unit distance corresponds to 500 km in the solar wind plasma frame. The title indicates the noise map used. The over-plotted straight white lines of different slopes indicated different
angles of attack.}
        \label{figure2}
 \end{figure*}

Whether single-point measurements, as with Parker Solar Probe, measure the parallel or perpendicular fluctuations to some local magnetic field, depends on the local solar wind speed, spacecraft speed, the variance, power, and spectral anisotropy of the perturbations, and the angle of the local magnetic field direction with respect to the spacecraft trajectory in the plasma frame. The total derivative in the spacecraft's frame of reference, equivalent to the derivative of spacecraft temporal data, is
\begin{equation}
    \frac{d}{dt} = ({V_{SC}}_R - {V_{SW}}_R)\frac{\partial}{\partial R} +
    ({V_{SC}}_T - {V_{SW}}_T)\frac{\partial}{\partial T} +  ({V_{SC}}_N - {V_{SW}}_N)\frac{\partial}{\partial N},
    \label{totalder}
\end{equation}
where $V_{SC}$ and $V_{SW}$ are the spacecraft speed and the solar wind flow speed, respectively. The triad $(R,T,N)$ represents the radial, tangential, normal coordinates, respectively, with $R$ pointing towards the spacecraft from the Sun's center, $T = \omega \times R/|\omega \times R|$ where $\omega$ is the Sun's spin axis, and $N$ completes the right-handed triad. One way of estimating how well one can measure parallel or perpendicular fluctuations is by calculating the cross-correlation between the total derivative and the component-wise spatial derivatives, as a function of angle of attack. The spatial derivative is proportional to the first-order structure function, defined by $S^1(\mathbf{r}) = \langle[\delta f(\mathbf{r})]\rangle$, where $\delta f(\mathbf{r}) = f(\mathbf{R}) - f(\mathbf{R+r})$, the difference of fluctuations at scale and direction $\mathbf{r}$. We again neglect variation in the normal direction. We calculate the cross-correlations in the following way. First, we calculate 2D matrices obtained by shifting the noise maps along both axes by varying amounts, and then subtracting the shifted matrix from the original. The purely parallel or perpendicular shifts are cross-correlated with matrices having both parallel and perpendicular shifts, which represents the total derivative. The cross-correlation is defined as:
\begin{equation}
    c_k = \sum_n a_{n+k} b_n,
\end{equation}
where $a$ and $b$ are the flattened matrices resulting in one-dimensional arrays, normalized such that the cross-correlation value is between $[0,1]$:
\begin{equation}
    a = \frac{a - \left<a\right>}{std(a) len(a)}, \quad
    b = \frac{b - \left<b\right>}{std(b)},
\end{equation}
where $<>$ denotes the arithmetic mean, $std(x) = \sqrt{\left<(x - \left<x\right>)^2\right>} $ is the standard deviation of the elements in $x$, and $len(x)$ is the total number of elements in the $x$ array. See Fig~\ref{figure2} for cross-correlations between total and partial derivatives for a noise map with a specific spectral and power anisotropy.
 We repeat this procedure for noise maps of different spectral and power anisotropies, the latter being defined at the highest wave number, representing the limit of the MHD inertial range in the solar wind, around 500 km, corresponding to the distance between two array elements in the noise maps. Investigating Fig.~\ref{figure2}, it is clear that for the specific spectral and power anisotropy chosen, perpendicular variations are better correlated than parallel variations with the total derivative, except for smaller angles of attack of around $15^\circ$ with respect to the magnetic field direction. Also note that the correlations are decreasing with increasing separation, both in the parallel and perpendicular direction, as the power anisotropy is scale-dependent when there is a spectral anisotropy, as in the current case. In the case of no spectral anisotropy ($k_\perp \sim k_\parallel \sim -5/3$), power anisotropy is scale-independent and there is no decreasing correlation with distance, resulting in a considerably stronger perpendicular correlation overall.

\begin{figure*}
    \centering
    \includegraphics[width=0.33\textwidth]{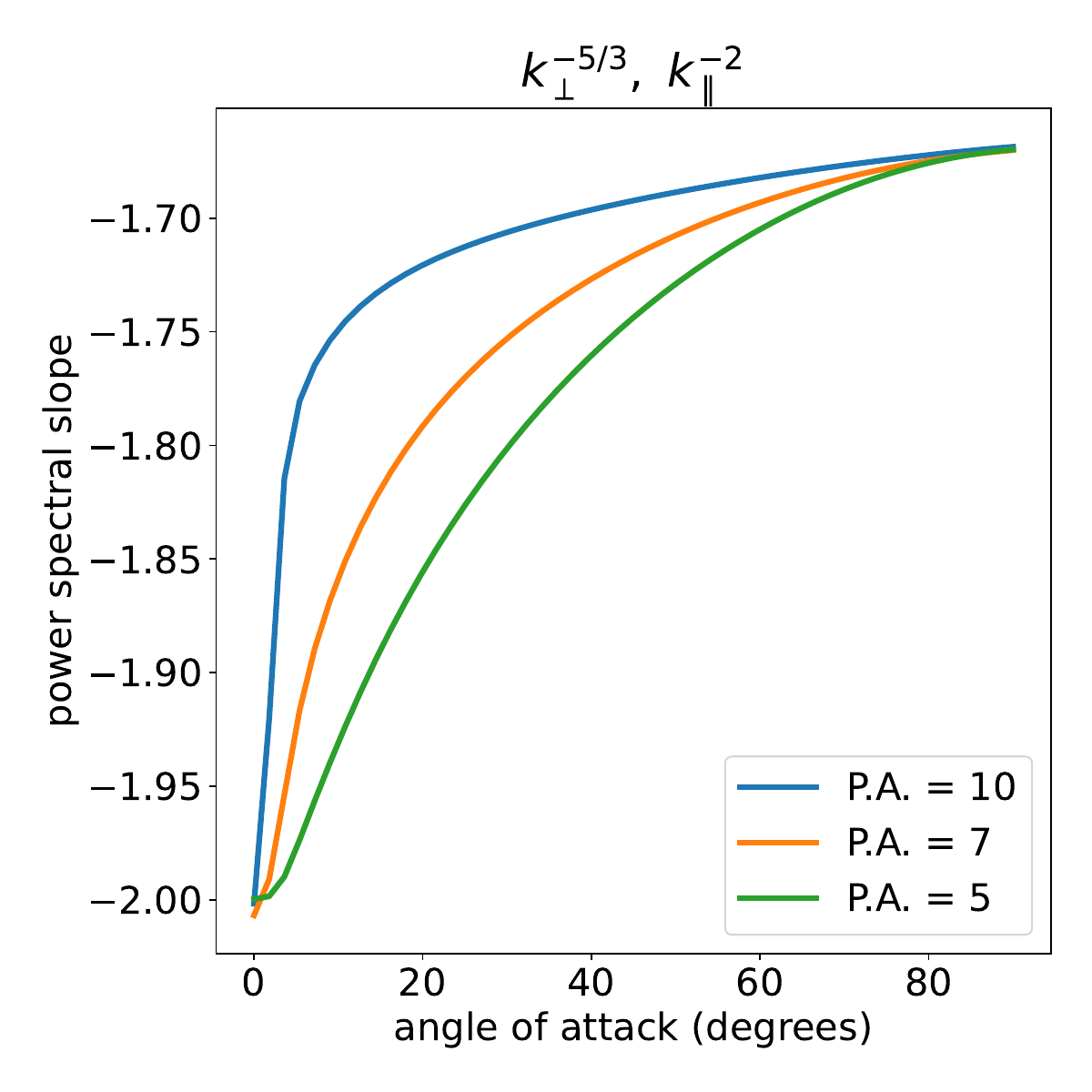}
    \caption{Slopes of average 1D power spectra as a function of the angle to the magnetic field direction, for the noise maps with spectral anisotropy, for different power anisotropies.}
    \label{figure4}
\end{figure*}

In Fig.~\ref{figure4}, the power spectral slopes of the two noise maps with spectral anisotropy and different power anisotropies are shown as a function of the angle to the magnetic field direction. These curves are obtained by applying a rotation transform to the noise map by specific angles in increments of $5^{\circ}$ from $0^{\circ}$ to $90^{\circ}$ , and then averaging 1D spectra along the axes. 
Fig.~\ref{figure4} can be compared to the similar plot of \citet{2008PhRvL.101q5005H}, derived from real solar wind data. The effect of power anisotropy on the variation of the slope with the angle is in agreement with the correlation plots: larger power anisotropy makes it more difficult to measure parallel variations. Interesting to note is that the true perpendicular spectrum is only attained for perpendicular trajectories, and there seems to be no constant plateau after some angle, contrary to what observational analysis suggests in \citet{2008PhRvL.101q5005H}. In the case of the noise maps, the transitional curve between parallel and perpendicular spectra is determined by the shape of the copula in Eq.~\ref{filter2}. Understanding the physical origins of such transitional curves determined from observational data might be revealing of physics of solar wind fluctuations.

\subsection{Application to PSP measurements}

\begin{figure*}
    \centering
    \includegraphics[width=0.5\textwidth]{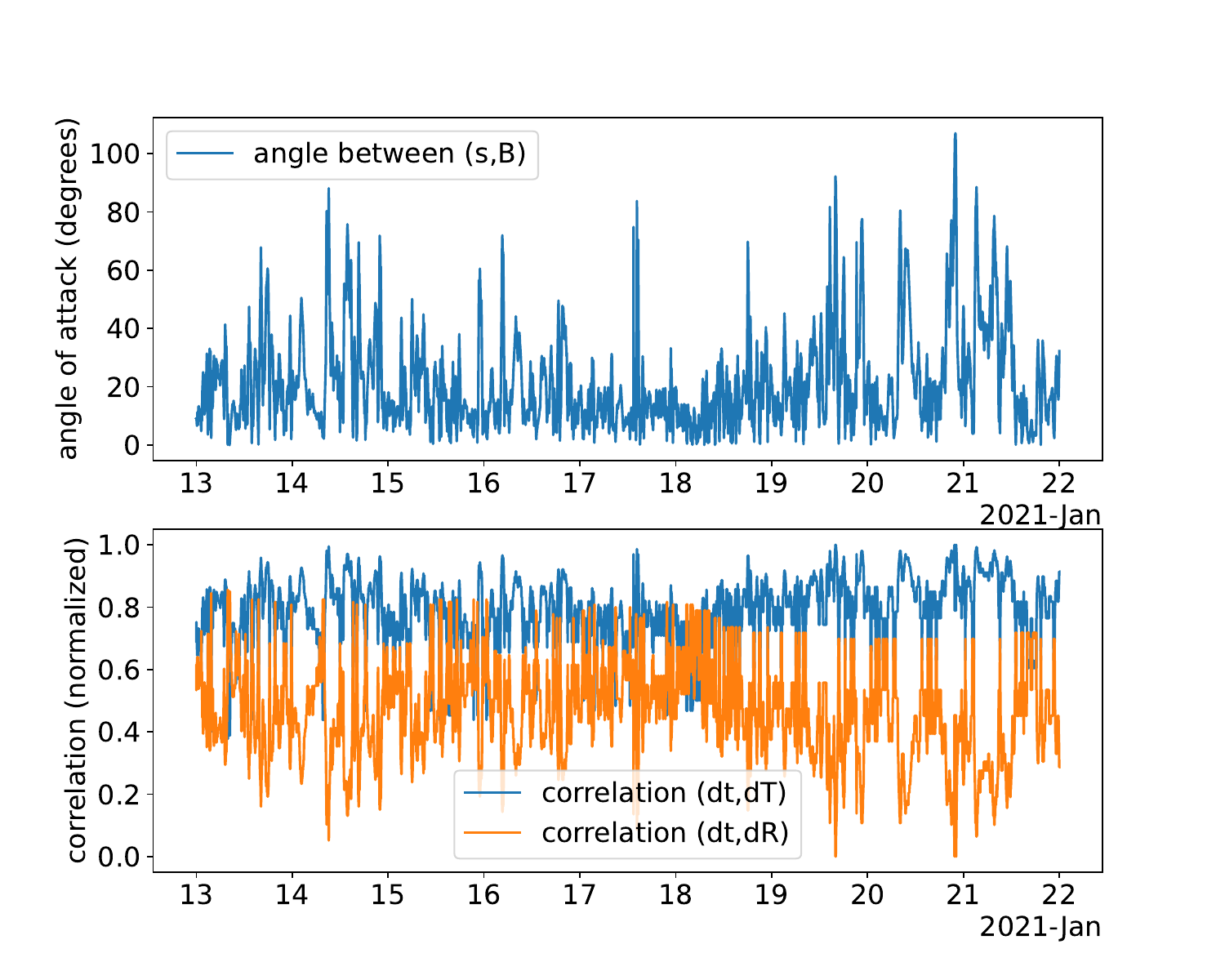}
    \caption{The measured angle of attack to the local mean magnetic field (top), and the corresponding normalized cross-correlation coefficient between the total and radial (parallel) or tangential (perpendicular) derivatives, respectively (bottom).}
    \label{figure7}
\end{figure*}

Based on the measured angle of attack values, by assuming a specific power and spectral anisotropy of the noise map that best matches the spectral properties of the solar wind, it is possible to make various predictions concerning the real data. For example, one can estimate the likelihood that measured perturbations represent parallel or perpendicular variations in the plasma frame, based on the cross-correlation maps in Fig.~\ref{figure2}. Here we mostly showcase the various analyses possible using this toolkit by employing a noise map with spectral anisotropy ($k_\perp \sim -5/3; k_\parallel \sim -2$), and a power anisotropy of 7. Thus, findings based on this noise map are only valid as long as the spectral properties of the noise map approximate well those of the solar wind. We do not claim that the chosen spectral and power anisotropy of the noise map is the representative or typical spectral and power anisotropy of the solar wind fluctuations for this particular PSP encounter. Caveats of choosing a single noise map for the whole duration of the encounter and of these particular values of spectral and power anisotropy will be discussed in the Conclusions section. 
Calculating the spatial distance in the plasma frame as the relative speed multiplied by the cadence, one can assign a normalized correlation value to each pair of measured angle of attack and distance values, shown in Fig.~\ref{figure7} for the noise map in Fig.~\ref{figure01}.

\begin{figure*}
    \centering
       \begin{tabular}{@{}cc@{}} 
        \includegraphics[width=0.33\textwidth]{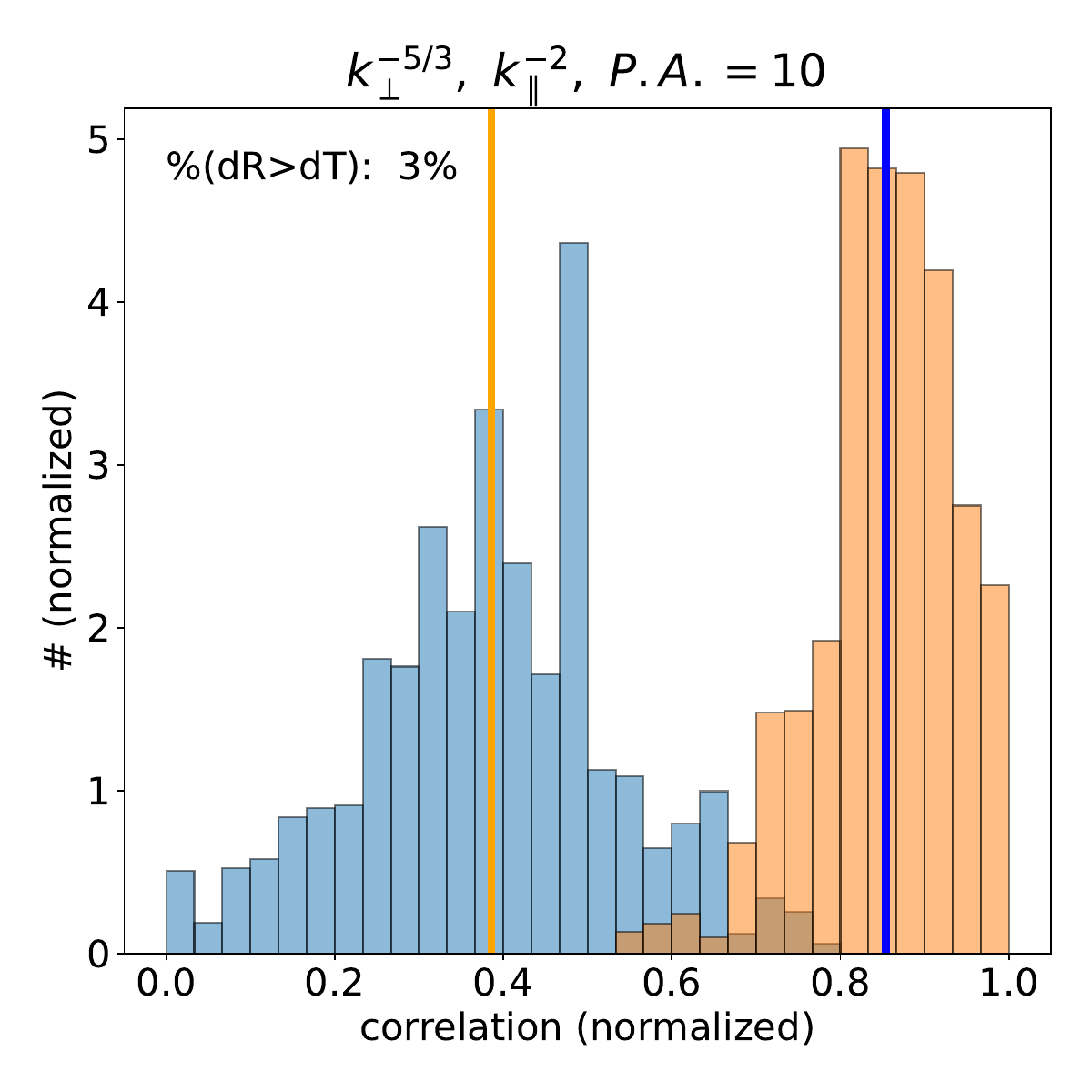}
        \includegraphics[width=0.33\textwidth]{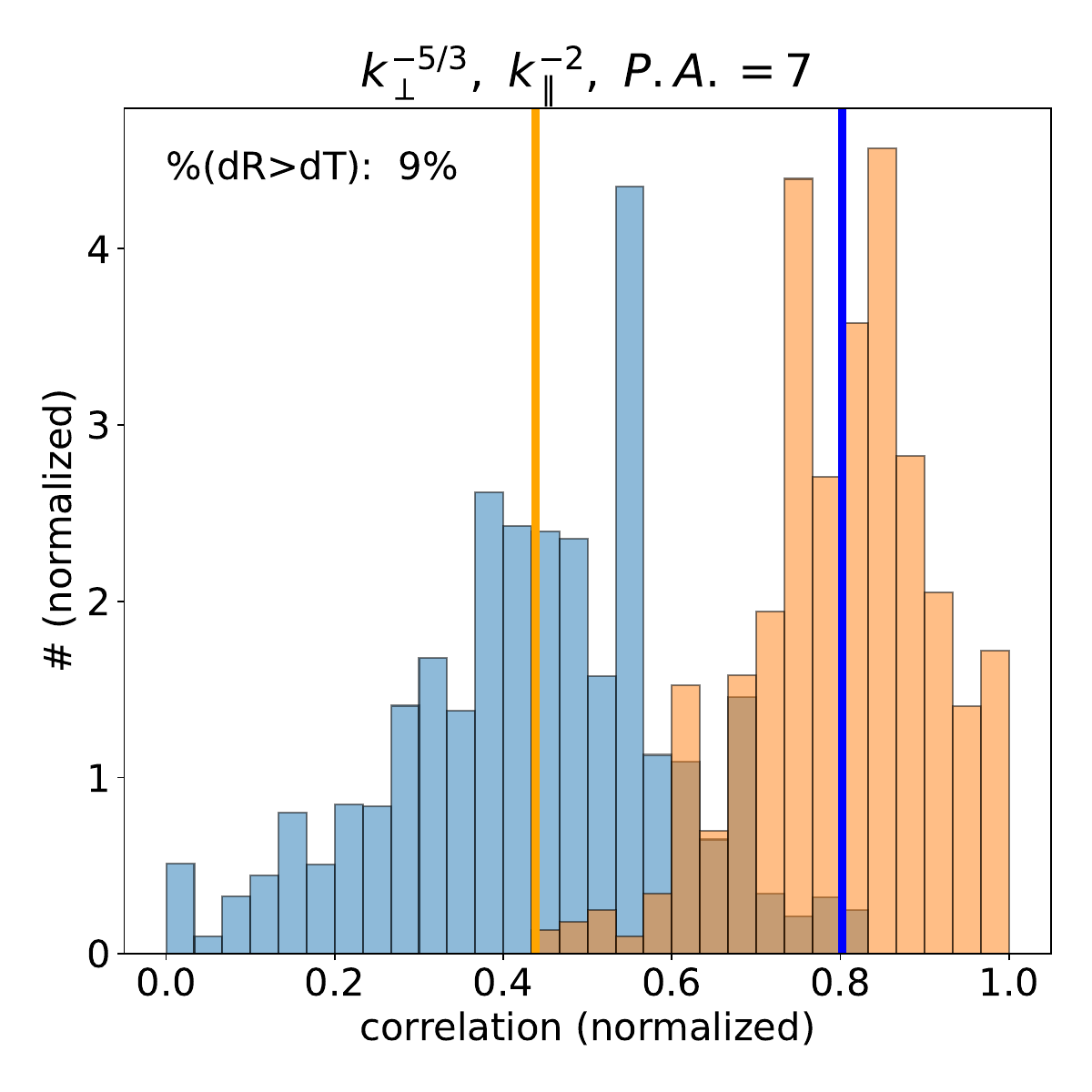} \\
        \includegraphics[width=0.33\textwidth]{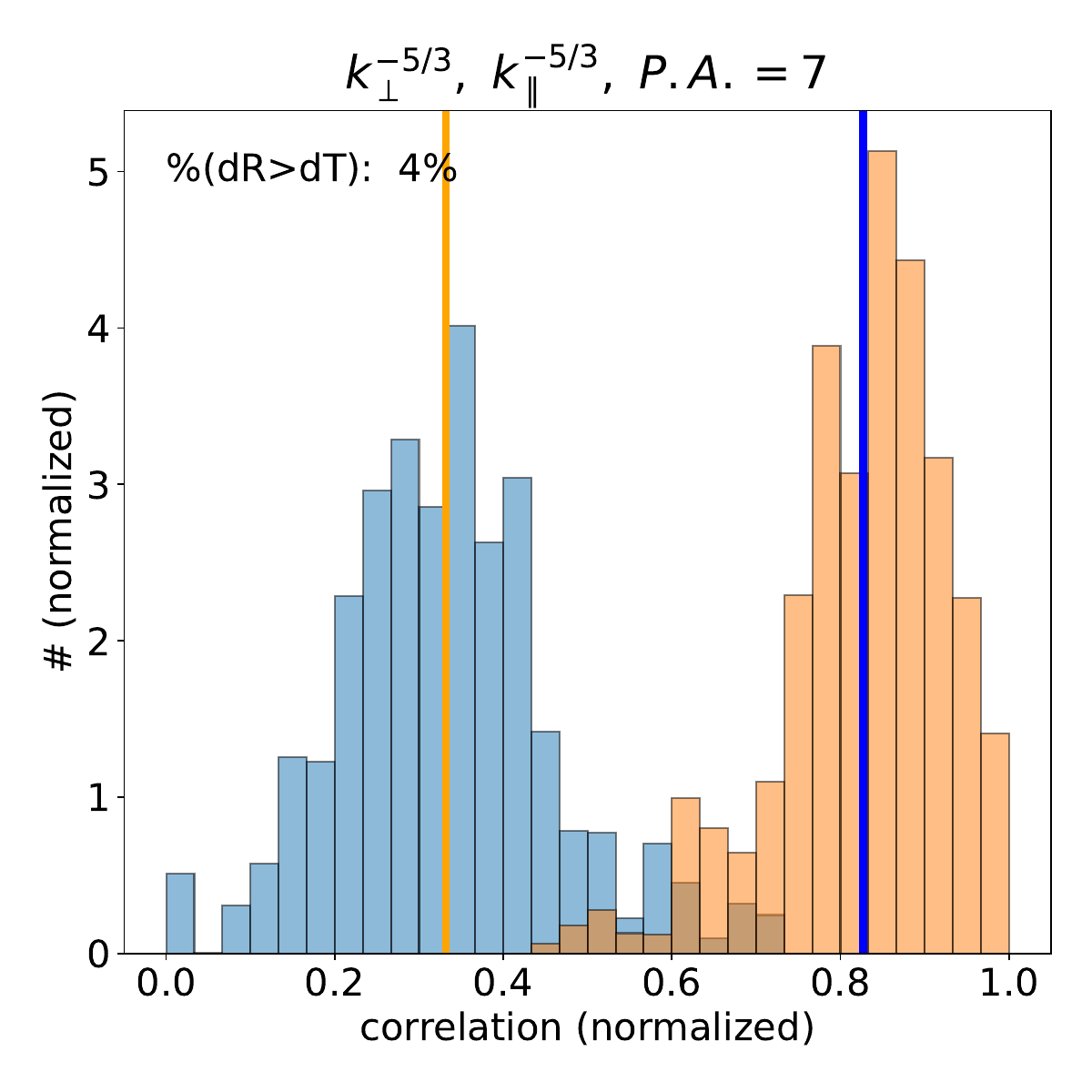} 
        \includegraphics[width=0.33\textwidth]{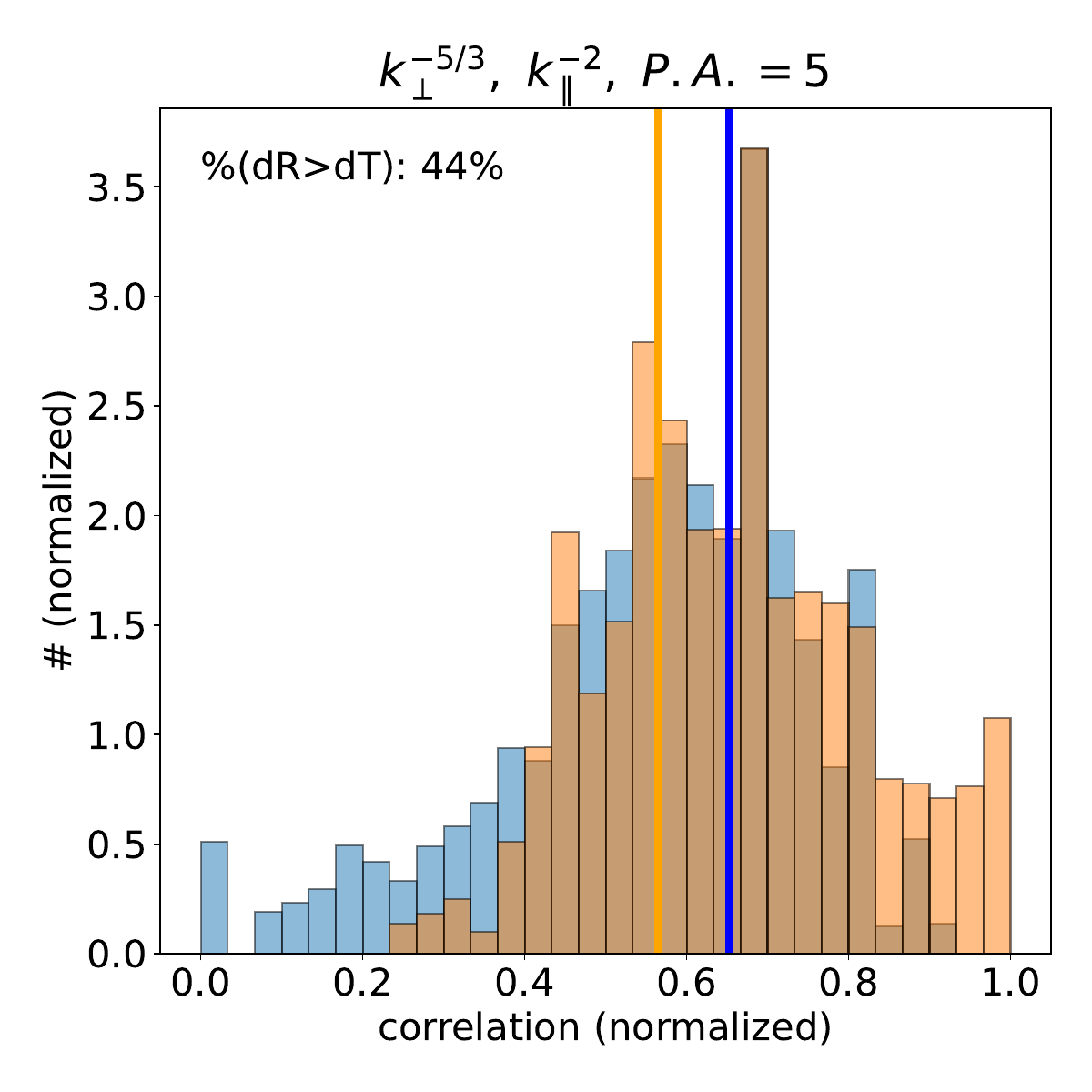} 
       \end{tabular}  
        \caption{Histograms with the distribution of correlation coefficients obtained by scanning through the correlation maps with the angles of attack obtained from PSP measurements, using a mean field with moving average of 30 min. The specific spectral and power anisotropies are shown at the top of each histogram plot. The orange (blue) bars represent the perpendicular (parallel) correlations, and the vertical lines represent the mean value of the correlation for the whole analyzed encounter period. Also shown is the percentage of points for which the parallel correlation value is higher than the perpendicular one.}
        \label{figure8}
\end{figure*}

The mean plasma frame distance between two measurements with the present cadence is 1350 km (minimum 740 km, maximum 2370 km), which corresponds to a mean power anisotropy of 6 (minimum 5, maximum 6.8) in the noise map of Fig.~\ref{figure01}. Looking at Fig.~\ref{figure7}, it is again clear that most of the temporal variation in the spacecraft frame is given by perpendicular variations in the plasma frame, for a noise map with spectral anisotropy and a power anisotropy of 7. The statistics of correlations are shown in Fig.~\ref{figure8} also for the other noise maps in Table~\ref{table1}.

The dominance of plasma frame perpendicular variations in the spacecraft frame temporal variation is increasing with increasing power anisotropy and with decreasing spectral anisotropy. As such, for a power anisotropy of 7, there is a dominance of perpendicular variations in the temporal variation, parallel variation having a better correlation for only 9\% of the time. For a power anisotropy of 5, the average correlation values with parallel and perpendicular variations are very close, meaning that in this case high correlations with either parallel or perpendicular variations are only achievable for nearly parallel or perpendicular angles of attack, respectively. 

\begin{figure*}
    \centering
       \begin{tabular}{@{}cc@{}} 
        \includegraphics[width=0.33\textwidth]{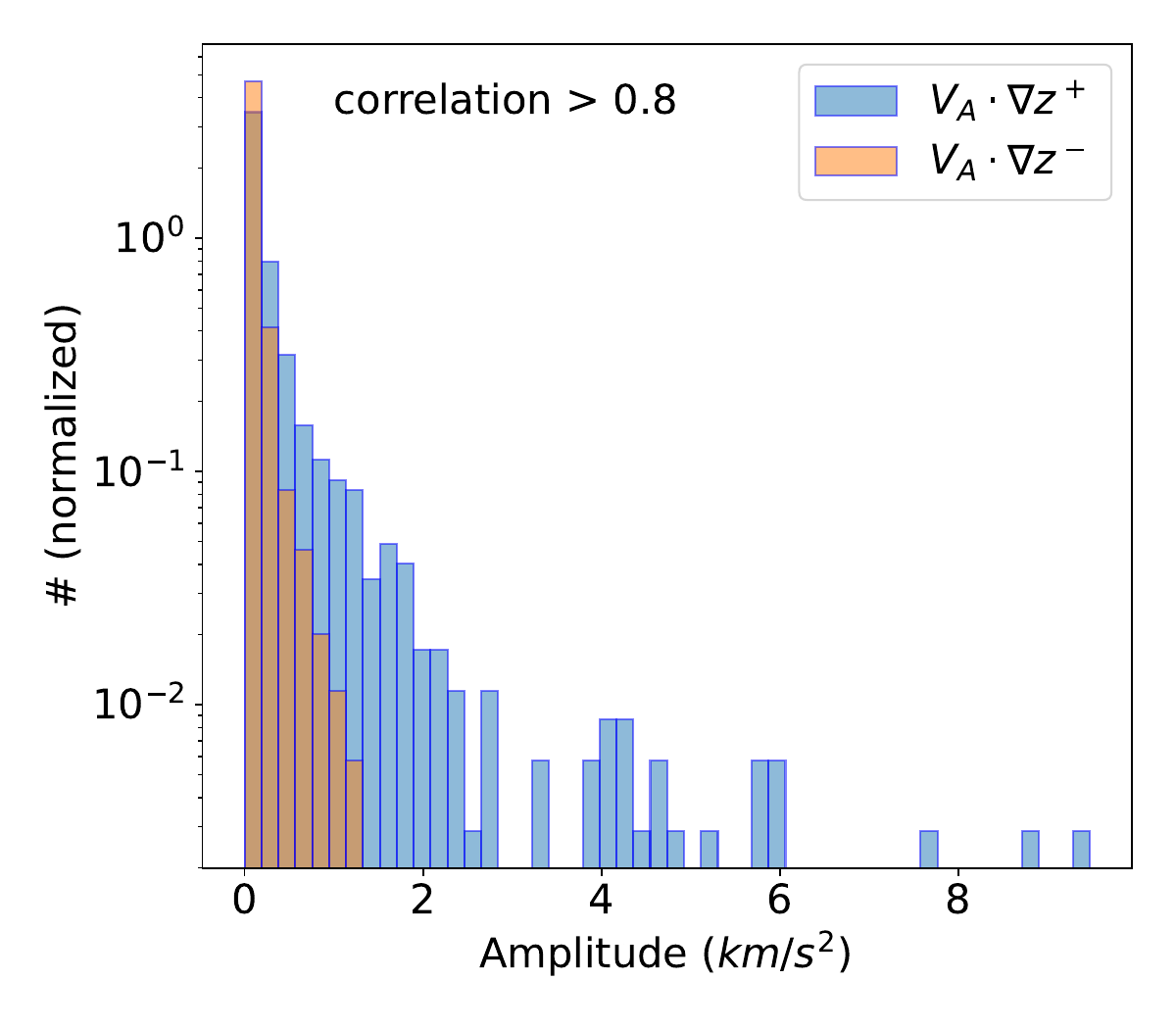}
        \includegraphics[width=0.33\textwidth]{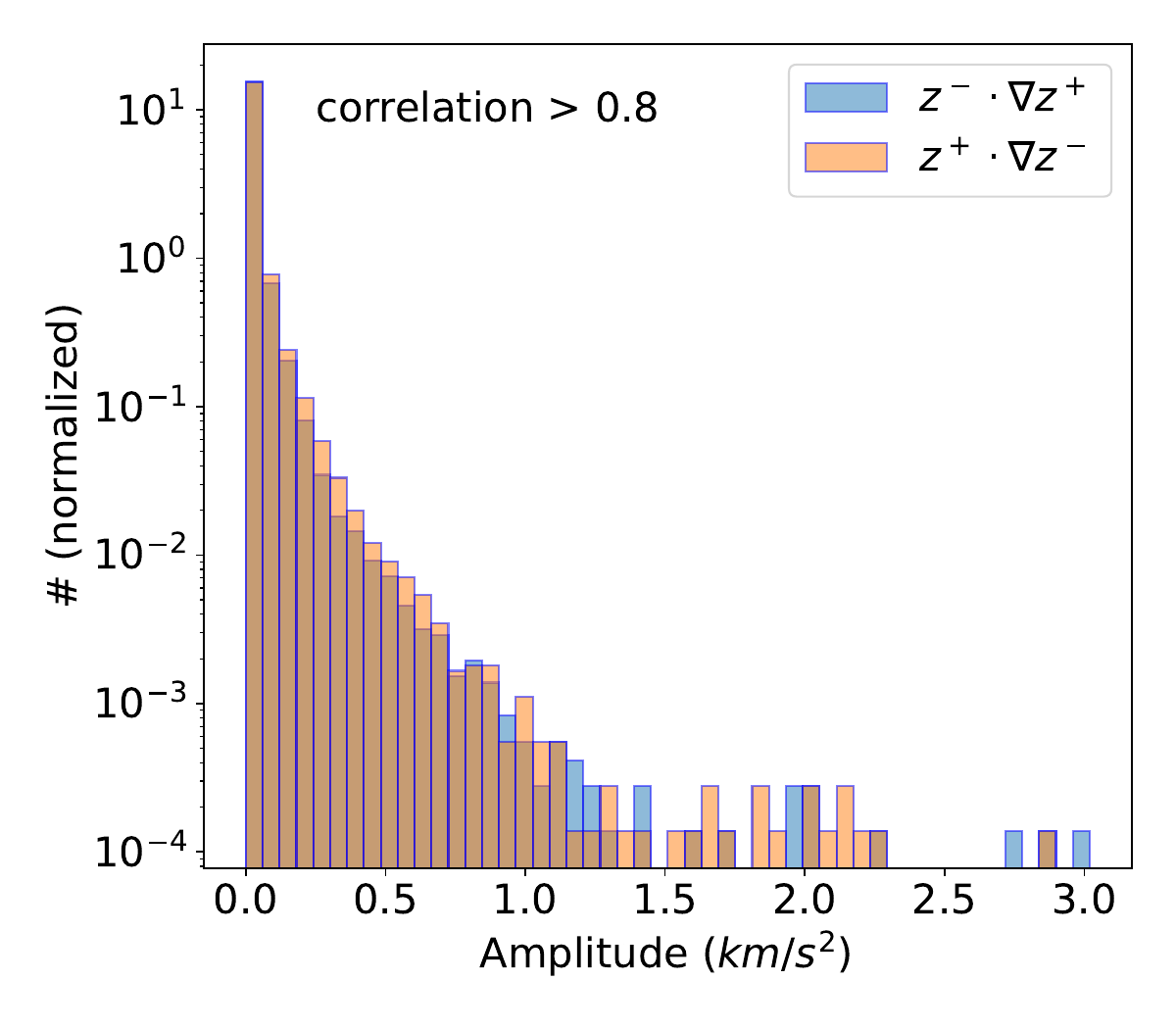} 
       \end{tabular}  
        \caption{Histograms showing the absolute value of the calculated linear advection (left) and nonlinear advection (right), for correlations above 0.8 with parallel or perpendicular variations, respectively. The calculated advection represents the component in the direction of the spacecraft velocity relative to the local plasma frame. Advection terms are calculated for both Els\"{a}sser variables.}
        \label{figure9}
\end{figure*}

Using the correlation values shown in Fig.~\ref{figure7} calculated specifically for the PSP measurements, it is possible to statistically study various plasma frame quantities derived from the PSP measurements, while being able to assign a value to how accurately these measurements represent parallel or perpendicular variations. Here we show the example of a linear and nonlinear advection study, based on data points for which either the perpendicular and parallel correlations are above 0.8, thus showing very good correlation. The noise map used is the same as in Fig.~\ref{figure01}, and thus the number of points satisfying the larger than 0.8 correlation criterion is much higher for perpendicular correlations than parallel ones. Still, there seem to be enough parallel correlation points ($\approx 5000$) for a statistical study. In Fig.~\ref{figure9}, the linear advection, measured for high parallel correlation, and nonlinear advection, measured for high perpendicular correlation, are shown. 
The perturbations are rotated to a reference frame pointing in the relative direction of the spacecraft velocity in the plasma frame, which is almost identical to the radial direction in the case of high parallel correlation. In this sense, the calculated advection represents a component of the total advection vector, with a weight of almost 100\% for linear advection and around 50\% for nonlinear advection. It is important to keep in mind that the linear and nonlinear advection calculated in this way are not for the same plasma element, as high parallel or perpendicular correlation regions do not overlap. The plots in Fig.~\ref{figure9} show that linear advection is higher for the dominant Els\"{a}sser variable (here $z^+$), as expected. The nonlinear advection for $z^\pm$ are of the same magnitude, implying that nonlinear advection is relatively more important for the minority Els\"{a}sser variable (here $z^-$). It appears that, by taking into account that the calculated nonlinear advection component is on average 50\% of the total value, the linear and nonlinear advection have around the same values for $z^-$, while the nonlinear advection being around half the linear advection for $z^+$. Thus, the advection of $z^-$ appears to be in critical balance, while $z^+$ could be in critical balance for short intervals, in a possibly intermittent manner. \par
\begin{figure*}
    \centering
    \includegraphics[width=0.5\textwidth]{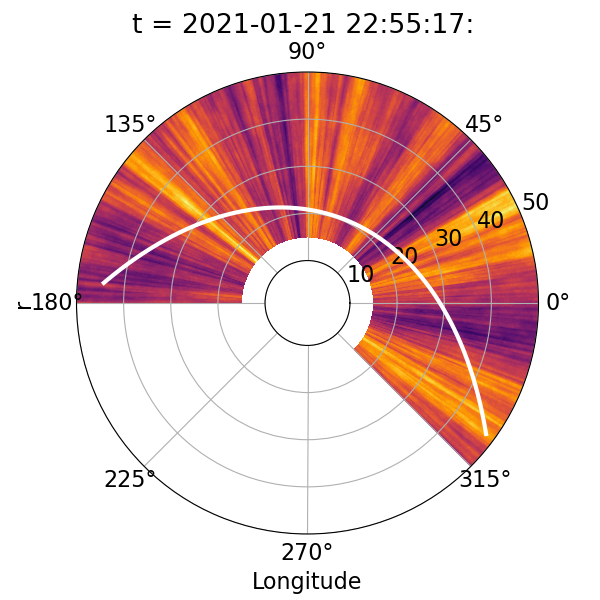}
    \caption{Polar plot showing the ecliptic plane in the HCI frame, with the PSP orbit during the 7\textsuperscript{th} encounter (here shown from the 13\textsuperscript{th} to 21\textsuperscript{st} of January 2021), the trajectory being depicted by a white curve. The background is given by the noise map in Fig.~\ref{figure01}. Radial distance in units of solar radii. An animation of this figure is available online.}
    \label{figure6}
\end{figure*}
\begin{figure*}
    \centering
    \includegraphics[width=0.5\textwidth]{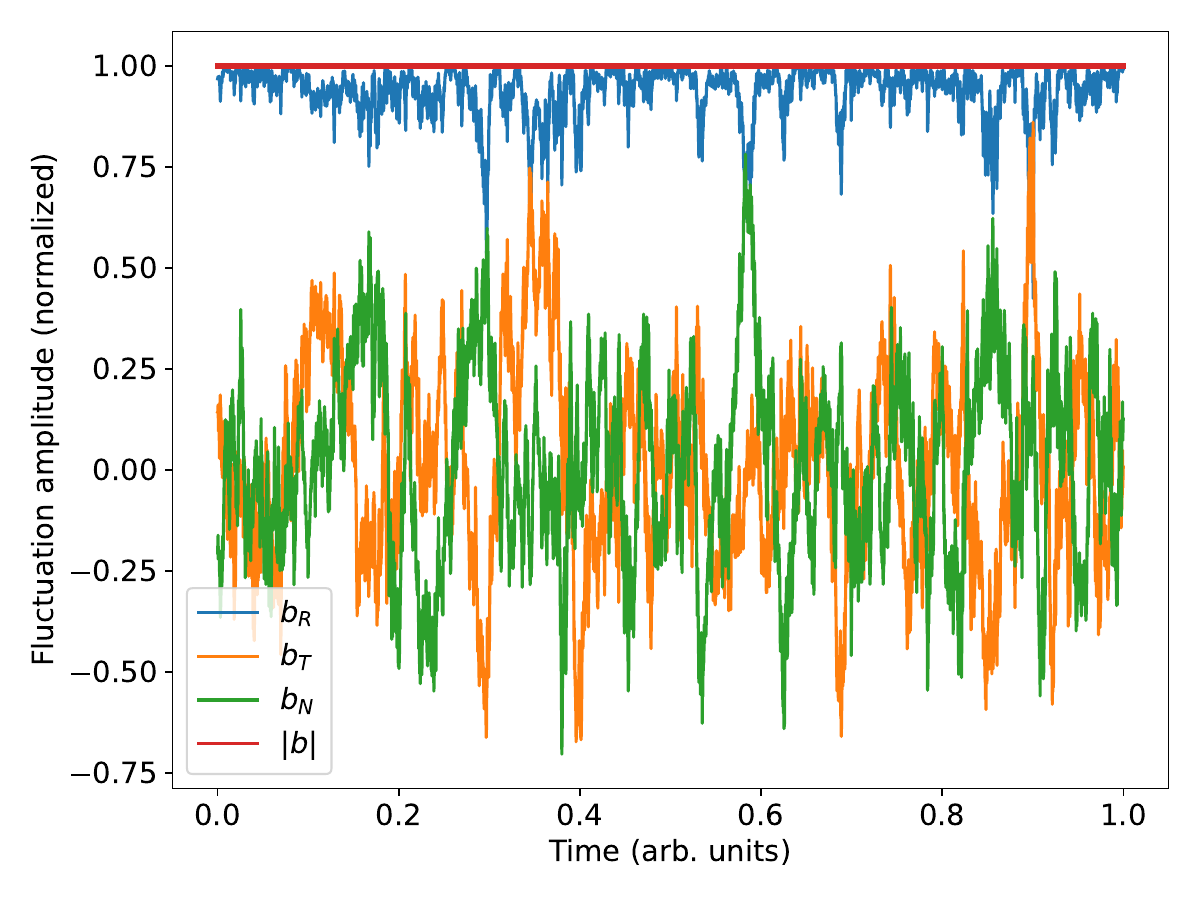}
    \caption{Synthetic magnetic field perturbations, derived from the noise map shown in Fig.~\ref{figure01}, showing each component and the total magnitude, which is constant.}
    \label{figure5}
\end{figure*}
\begin{figure*}
    \centering
    \includegraphics[width=0.5\textwidth]{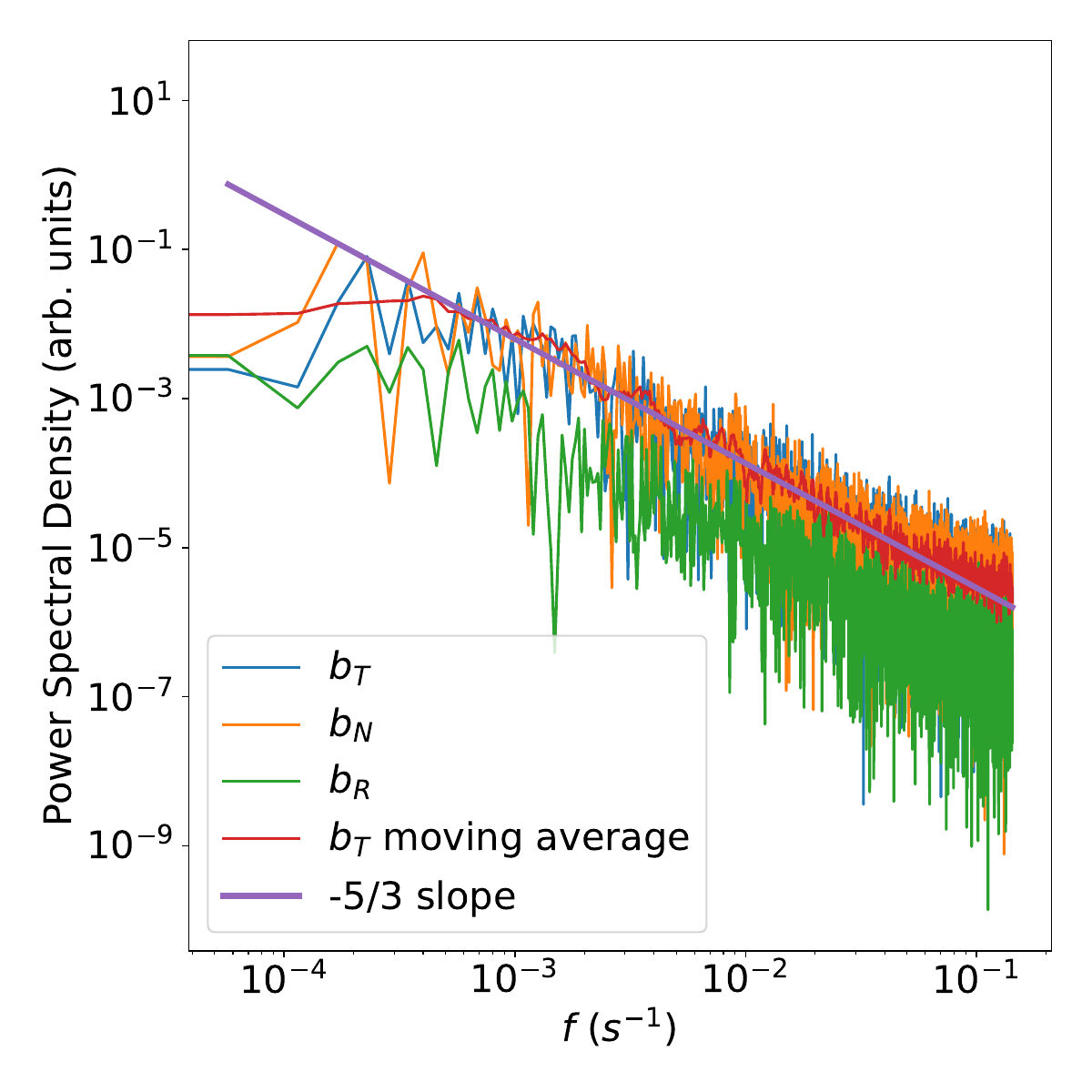}
    \caption{Power spectral density for the three synthetic components of the magnetic field, for the whole duration of the analyzed PSP encounter. Also shown is a moving average of the tangential component, to be more easily compared to the overplotted $-5/3$ slope.}
    \label{figure12}
\end{figure*}
\begin{figure*}
    \centering
    \includegraphics[width=0.5\textwidth]{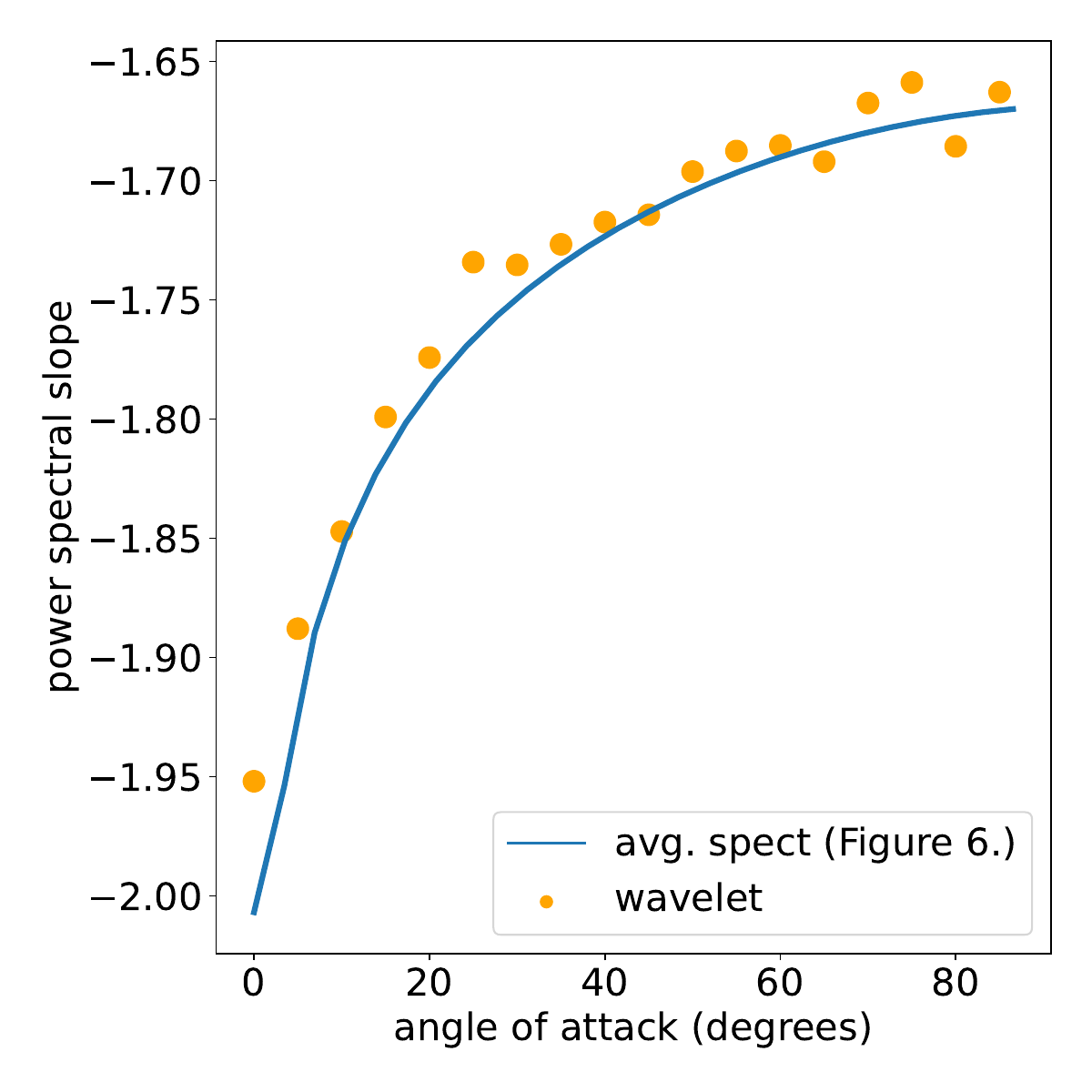}
    \caption{Angle of attack-dependent spectra, calculated in two different ways: by averaging over the spectra of 1D lines at a specific angle to the noise map principal axes (also shown in Fig.~\ref{figure4}), and by applying the wavelet technique as described in \cite{2008PhRvL.101q5005H} to the synthetic 1D data.}
    \label{figure13}
\end{figure*}

Another example of how synthetic data can be of help in understanding one-point spacecraft measurements is the following. It is possible to extract synthetic perturbations through the noise map by simulating a PSP fly-through of the noise map, with background wind speed, spacecraft speed, and angle of attack given by PSP measurements. This is shown in Fig.~\ref{figure6}.
The angle of attack variation is simulated by multiplying the relative velocity components of the spacecraft with the cosine and sine of the measured angle values. Then, synthetic measurements can be obtained by extracting data points from the noise map along the simulated PSP trajectory. The generation of synthetic velocity or magnetic field perturbation vectors with three components is possible, by using a different noise map for each component. For example, magnetic field perturbations with constant magnitude, that is, zero compressibility or spherical polarization, can be constructed by generating two different random noise maps for, e.g., the tangential and normal components, and defining the radial component as: 
\begin{equation}
b_R(r_\parallel,r_\perp) = \sqrt{max(b_T^2+b_N^2) - b_T^2 - b_N^2},  
\end{equation}
where the $max()$ finds the maximal value within a specific array. Note that in this case the radial magnetic field is always positive, thus no synthetic magnetic switchbacks that flip the magnetic field polarization are present. An example of synthetic magnetic field perturbation time series constructed in such a way is shown in Fig.~\ref{figure5}.
Using such synthetic 1D measurements, the analysis presented earlier can be repeated, such as the correlation between the gradient of the synthetic 1D measurement and the spatial gradients at the location of the spacecraft. Correlations derived in such way show a very good agreement with the correlations derived directly from correlation maps (Fig.~\ref{figure7}), with some fluctuations that change for each set of random numbers used to generate the noise map. The power spectrum of the synthetic 1D fluctuations can also be analyzed, with the power spectrum for the total duration and the noise map in Fig.~\ref{figure01} shown in Fig~\ref{figure12}.
The variance anisotropy (difference in amplitude between parallel and perpendicular components of the synthetic perturbations) is around 15, in the upper range of previously reported variance anisotropies in real data, with the typical value being around 6 \citep[e.g.,][]{2006JGRA..111.9111S}, and 9 as reported by \citet{1971JGR....76.3534B}. The measured total power spectral slope is close to $-5/3$, indicating yet again the dominance of the perpendicular variations in the temporal data. We have conducted a wavelet analysis as well in order to study the angle of attack-dependent power spectra of the synthetic data, with similar parameters to the one presented in \citet{2008PhRvL.101q5005H}, and the result is shown in Fig.~\ref{figure13}.
The wavelet technique applied to the synthetic 1D data shows a very good agreement with the one derived through rotating the noise map and then averaging the spectra over all 1D lines, also shown Fig.~\ref{figure4}. It would be revealing in the future to perform all other analysis techniques that were used to disentangle parallel vs. perpendicular spectra, e.g., Hilbert spectral analysis \citep{2019ApJ...887..160T}, to see if different results are obtained. \par
\begin{figure*}
    \centering
    \includegraphics[width=0.5\textwidth]{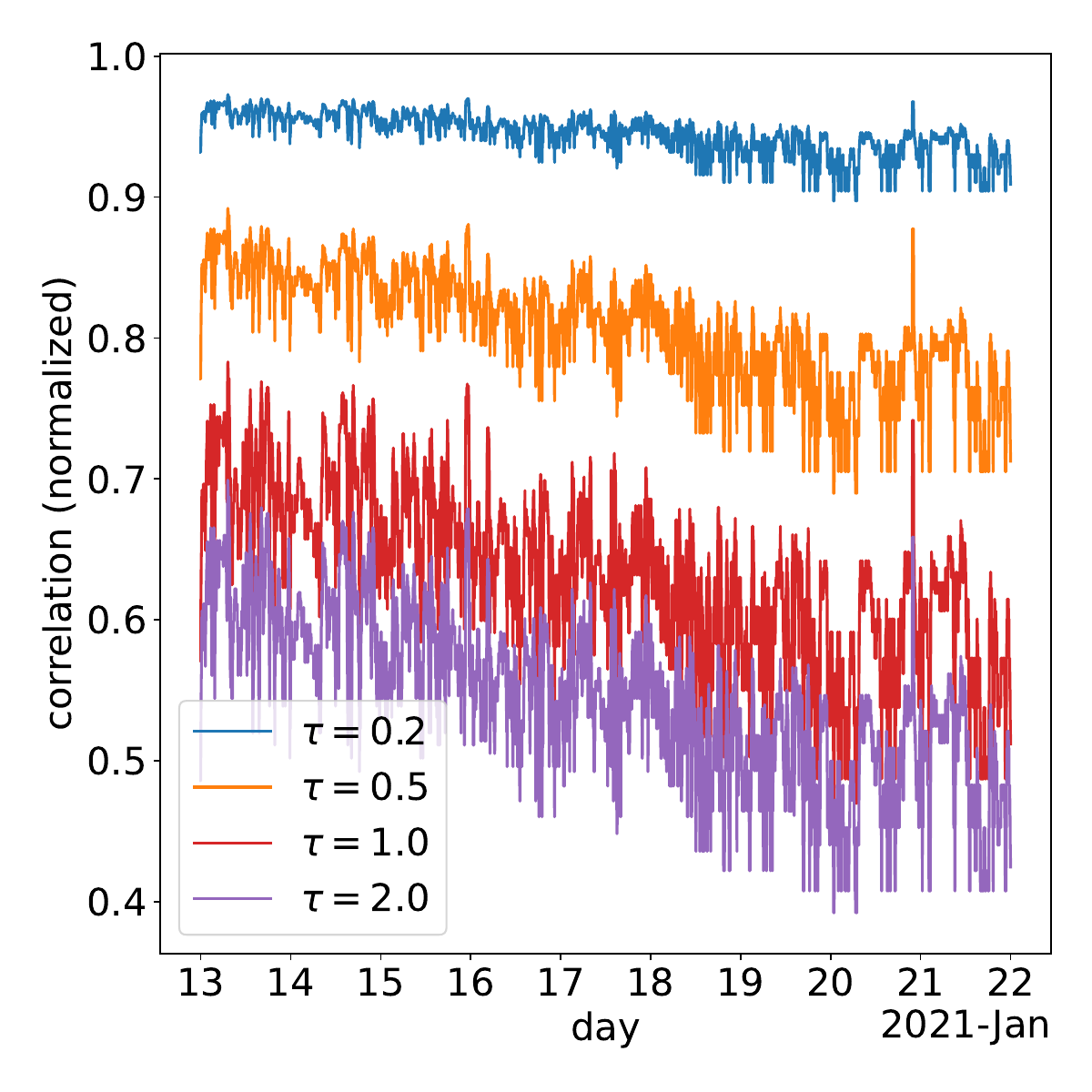}
    \caption{Correlation between the 1D synthetic data extracted from an unchanging `frozen-in' noise map and the noise map undergoing deformations of a magnitude set by $\tau$, for the whole analyzed period of the encounter.}
    \label{figure11}
\end{figure*}

As a last example of applicability of synthetic maps, one could also use the synthetic data to determine the limit of validity of Taylor's hypothesis, by considering local deformations of the noise map along the followed trajectory. Then, the unaltered signal (i.e., fluctuations `frozen-in') can be correlated with the signal suffering deformations proportional to some deformation constant $\tau$. The results are shown in Fig.~\ref{figure11}.
The deformation constant can be interpreted as the inverse of some deformation time normalized to the cadence, with its value reflecting the relative amount of advection or deformation suffered by fluctuations in the plasma frame, with a value of $\tau = 1$ representing deformation on the order of the fluctuations. This can be taken to be equivalent to the ratio of the velocity fluctuation amplitude in the plasma rest frame ($\delta v$) to the spacecraft speed relative to this frame ($V_\perp$), in the perpendicular direction to the magnetic field, $\tau = \delta v/V_\perp$. In this sense, $\tau$ is proportional to the quantity $\epsilon$ used in \citet{2021A&A...650A..22P}. Note that using $\epsilon$ as defined in \citet{2021A&A...650A..22P} here would lead to highly varying `deformation' rates as the angle of attack is highly varying. Instead, for demonstrative purposes, we opted for a steady or average rate of deformation over time as quantified by the value of $\tau$. Thus $\tau$ here can be interpreted as an average $\epsilon$ multiplied by $\sqrt{2}$. The deformations are introduced by Fourier transforming a square region around the measurement point in the noise map (10 pixels, or 5000 km wide), and rotating (phase shifting) the resulting complex $k$-plane by a different random amount at each discrete ($k_\perp,k_\parallel$). The maximum allowed rotation is then given by $\tau$, in units of a full rotation ($2\pi$). Note that the random numbers scale as $k$, as the deformation time is proportional to the scale of the fluctuations. Thus higher frequencies are deformed faster. The $k=0$ mode is not rotated, thus there is no change in mean value in the deformed square. The square is then inverse Fourier-transformed to obtain the locally deformed state of the noise map.
There are two main observations to be made by looking at Fig.~\ref{figure11}. First, clearly the correlation decreases as the deformation time decreases, as $\tau$ goes up. Second, the impact of the background flow speed is well visible, the correlation being higher for higher wind speeds before perihelion (13t\textsuperscript{th}-17\textsuperscript{th} Jan, see background wind speed in Fig.~\ref{figure1}). While there is still some correlation remaining for $\tau = 1$, that is, deformation on the order of the amplitude of the fluctuations, this is mostly due to the larger scales which remain better correlated. Thus, besides the wind speed, the linear and nonlinear deformation time has a strong influence on the validity of Taylor's hypothesis. The impact of the deformation on the power spectrum was also checked, and we found no impact on the overall spectra. This is understandable, as the deformations only impacted the phase of the fluctuations.

\section{Conclusions} \label{sec:concl}

Most of our understanding of solar wind dynamics comes from single spacecraft measurements, which provide magnetic field or particle data, or both, at some specified temporal cadence. Parker Solar Probe is one recent example of such a spacecraft. However, the solar wind is a three-dimensional, anisotropic plasma flow displaying strong fluctuations, with a mean magnetic field strongly varying in direction spatially, and with different properties along and across the mean field, a fact also predicted by several theoretical studies. In this sense, it is a great challenge to transform the temporal data of the spacecraft into the spatial plasma frame that is better suited to theoretical scrutiny. Even with Taylor's hypothesis being valid, separating between parallel and perpendicular spatial variations is not straightforward. Even more worrisome, in regions scanned by PSP closer to the Sun, the validity of Taylor's hypothesis is questionable. \par 
Initially motivated by attempting to determine the nonlinear advection term from solar wind measurements, we have devised a tool for approximating the likelihood that temporal measurements represent either parallel or perpendicular variations in the plasma frame, with the help of synthetic, anisotropic 2D noise maps. These noise maps can be constructed with specific power and spectral anisotropies, ideally ones that match with the studied solar wind period. Although determining the power and spectral anisotropies of solar wind data is a challenge in the first place, and thought to depend on a number of solar wind parameters, average representative values can be used for this purpose. More specifically, there is an ongoing debate under which conditions do fluctuations parallel to the local mean magnetic field scale as $k^{-2}$ or $k^{-5/3}$. Therefore we have analyzed multiple noise maps with different power and spectral anisotropies. \par 
The likelihood that the temporal variations represent parallel or perpendicular variations in the plasma frame is determined by calculating the first order structure functions (proportional to spatial gradients) from the noise maps, and then correlating these with either their perpendicular or parallel components. These correlation maps then assign a correlation coefficient, normalized between 0 and 1, between the total derivative and parallel or perpendicular derivatives, depending on spatial distance between two measurements and the angle between the local mean magnetic field and the relative velocity vector of the spacecraft in the plasma frame, referred to as `angle of attack'. \par 
Then, given real measurements of plasma frame distance between two data points, calculated from relative speed multiplied by the cadence, and angle of attack, calculated as the angle between the local mean magnetic field and the relative velocity vector of the spacecraft, we can assign perpendicular and parallel correlation values to each measurement point, which can be used to select intervals with likely high perpendicular or parallel correlation. 
The immediate outcome of this study is that power anisotropy, and to a lesser degree spectral anisotropy, have a strong effect on whether the measured temporal data comes from perpendicular or parallel variations in the plasma frame. The most important role is played by power anisotropy: a larger anisotropy leads to higher correlation between the temporal and perpendicular variations. Spectral anisotropy renders power anisotropy scale-dependent, decreasing it at larger scales, therefore indirectly having an impact on the correlations. For a noise map with spectral anisotropy of $k_\parallel \sim k_\perp^{2/3}$, and power anisotropy at the high frequency end of the inertial range of 7, most of the temporal variation is given by perpendicular variations in the noise map. Good correlation ($\geq 0.8$) with parallel variations is only encountered in 9\% of the total duration of E7. Without spectral anisotropy, the same power anisotropy would have only 4\% of the time good parallel correlation. \par
Some other applications of the synthetic maps in aiding data analysis were presented. Linear and nonlinear advection were calculated for intervals with good parallel and perpendicular correlations, respectively. These might indicate the advection of the minority Els\"{a}sser variable being critically balanced, while the majority Els\"{a}sser variable possibly only satisfies critical balance intermittently, being on average half the value of linear advection. \par 
Synthetic measurements can also be constructed for multiple components of perturbations based on the noise maps, including ones with spherical polarization or constant perturbation vector magnitude. These measurements are extracted from the noise map along the same trajectory (distance and angle of attack) as measured by PSP. The synthetic measurements obtained in such a way can be subjected to the same analysis methods as real data, with the advantage of also knowing their spatial variation. As such, we have obtained power spectra for the whole encounter duration of the synthetic data, showing a spectral slope close to $-5/3$. In order to reveal the parallel scaling, we have applied the same wavelet analysis as employed by \citet{2008PhRvL.101q5005H} to the synthetic data, revealing the angle of attack-dependent spectra of the noise map for the trajectory, which agrees well with the angle of attack dependent spectral slope obtained directly from averaging the noise map spectra. Synthetic measurements of this kind could be a tool to test and validate analysis methods applied to real data. Finally, we have tested Taylor's hypothesis by inducing localized deformations of the noise map along the relative spacecraft trajectory. Correlating the synthetic measurement from the non-deformed (static) noise map with the one from the deformed noise map, an upper limit of the linear or nonlinear timescale can be determined, for which Taylor's hypothesis is still valid. The correlation increases with wind speed, as expected. The correlation is still good ($\geq 0.8$) for the 7th encounter up to a relative deformation of $\tau = 0.5$ amplitudes at the smallest scales, within one cadence duration. \par 
There are a number of caveats of this study. First of all, the validity of various analyses based on the noise map as presented above are directly tied to how well the spectral and power anisotropy of the chosen noise map approximates that of the solar wind. As these properties of the solar wind are highly variable \citep[e.g.,][]{2022ApJ...924L...5Z}, care should be taken when choosing any single noise map to be representative for the whole interval analyzed. Here we have chosen the same noise map for the whole interval with specific spectral and power anisotropies in order to demonstrate the capabilities of this toolkit. Additionally, the dimensionality of the noise map could be made 3D for a fully replicated trajectory, and to further allow for asymmetry between the two perpendicular directions. Furhermore, the dimensions of the noise map could be increased, which would allow for the extraction of more synthetic data points, increasing the statistical significance of the performed analysis. Second, the noise maps used do not display intermittency as turbulence in the solar wind does. Intermittency is know to be very relevant for the determination of power and spectral anisotropy in solar wind measurements \citep{2014ApJ...783L...9W}. This could be improved by using random distributions with some specified kurtosis when generating the noise maps. The results of full 3D MHD simulations, which result self-consistently in anisotropic and intermittent turbulent fluctuations, have been used previously for such purposes. However, in the case of simulation data, the power and spectral anisotropies are not configurable but are the result of the dynamics, therefore no parametric studies on the impact of these properties is possible. Additionally, the obtained synthetic measurements do not display the $1/r^2$ decrease in average magnetic field magnitude with distance from the Sun, or heliospheric current sheet crossings. These could be included by modulating the fluctuations by the measured large-scale mean magnetic field. \par 
A necessary future extension of this study would be the constraining and updating of noise map properties by the measured average properties of the wind that is being analyzed. For example, in \citet{2022ApJ...924L...5Z,2023ApJ...951..141S}, it is shown that both the measured power and spectral anisotropy change with distance from the Sun, with spectral anisotropy being even scale-dependent, as also noted recently by \citet{2022FrASS...9.7393T,2023ApJ...947...45W}. Updated noise maps should take all these observations into account. Additionally, the influence of the violation of Taylor's hypothesis on the anisotropy measurements could also be thoroughly tested in the future by deforming noise maps as presented here. \par 
In summary, by generating and analyzing synthetic data based on noise maps with properties mimicking the properties of turbulence in the solar wind, we can gain better insights into the interplay between the temporal measurements in the spacecraft frame and spatial variations in the plasma frame, which allow for a better comparison between observations and turbulence theories. \\

\begin{acknowledgements} N.M. acknowledges Research Foundation – Flanders (Fonds voor Wetenschappelijk Onderzoek - Vlaanderen) for their support through a Postdoctoral Fellowship. This research was supported by the International Space Science Institute (ISSI) in Bern, through ISSI International Team project \#560: “Turbulence at the Edge of the Solar Corona: Constraining Available Theories Using the Latest Parker Solar Probe Measurements”. JV acknowledges support from NASA PSP-GI 80NSSC23K0208. TVD was supported by the European Research Council (ERC) under the European Union's Horizon 2020 research and innovation programme (grant agreement No 724326) and the C1 grant TRACEspace of Internal Funds KU Leuven. TVD has benefited from the funding of the FWO Vlaanderen through a Senior Research Project (G088021N) \end{acknowledgements}

\bibliography{Biblio}{}
\bibliographystyle{aa}

\end{document}